\documentclass[lettersize,journal]{IEEEtran}
\usepackage{amsmath,amsfonts}
\usepackage{algorithmic}
\usepackage{algorithm}
\usepackage{array}
\usepackage[caption=false,font=footnotesize,labelfont=rm,textfont=rm]{subfig}
\usepackage{textcomp}
\usepackage{stfloats}
\usepackage{url}
\usepackage{verbatim}
\usepackage{graphicx}
\usepackage{tabularx}
\usepackage{multirow}
\usepackage{cite}
\usepackage{amssymb}
\usepackage{bm}
\usepackage{epsfig,epsf,color,balance,cite}

\usepackage{amsmath}
\allowdisplaybreaks[4]

\begin{document}

\title{Energy-Efficient UAV-Enabled MEC Systems: NOMA, FDMA, or TDMA Offloading?}

\author{Qingjie Wu, Miao Cui, Guangchi Zhang, Beixiong Zheng, \IEEEmembership{Senior Member, IEEE},\\ Xiaoli Chu, \IEEEmembership{Senior Member, IEEE}, and Qingqing Wu, \IEEEmembership{Senior Member, IEEE}

\thanks{The work of Miao Cui and Guangchi Zhang was supported in part by Guangdong Basic and Applied Basic Research Foundation under Grant 2023A1515011980.
The work of Beixiong Zheng was supported in part by the National Natural Science Foundation of China under 62571193, Grant 62201214, Grant 62331022, and Grant 62301171, the Natural Science Foundation of Guangdong Province under Grant 2023A1515011753, Grant 2024A1515010013, and Grant 2022A1515110484, the Guangdong program under Grant 2023QN10X446 and Grant 2023ZT10X148, the GJYC program of Guangzhou under Grant 2024D01J0079 and Grant 2024D03J0006, and the Fundamental Research Funds for the Central Universities under Grant 2024ZYGXZR087.
The work of Qingqing Wu was supported by NSFC 62371289 and Shanghai JXW Fund JJ-GGFWPT-01-24-0030. \textit{(Corresponding authors: Miao Cui; Guangchi Zhang.)}}
\thanks{Qingjie Wu was with the School of Information Engineering, Guangdong University of Technology, Guangzhou 510006, China, and is currently with the School of Microelectronics, South China University of Technology, Guangzhou 511442, China (e-mail: qingjiewu55@163.com).}   
\thanks{Miao Cui and Guangchi Zhang are with the School of Information Engineering, Guangdong University of Technology, Guangzhou 510006, China (e-mail: cuimiao@gdut.edu.cn; gczhang@gdut.edu.cn).}
\thanks{Beixiong Zheng is with the School of Microelectronics, South China University of Technology, Guangzhou 511442, China (e-mail: bxzheng@scut.edu.cn).}
\thanks{Xiaoli Chu is with the School of Electrical and Electronic Engineering, The University of Sheffield, S1 3JD Sheffield, U.K. (e-mail: x.chu@sheffield.ac.uk).}
\thanks{Qingqing Wu is with the Department of Electronic Engineering, Shanghai Jiao Tong University, Shanghai 200240, China (e-mail: qingqingwu@sjtu.edu.cn).}
\thanks{Corresponding authors: Miao Cui and Guangchi Zhang.}
}



\maketitle

\begin{abstract}
Unmanned aerial vehicle (UAV)-enabled mobile edge computing (MEC) systems can use different multiple access schemes to coordinate multi-user task offloading. However, it is still unknown which scheme is the most energy-efficient, especially when the offloading blocklength is finite. To answer this question, this paper minimizes and compares the MEC-related energy consumption of non-orthogonal multiple access (NOMA), frequency division multiple access (FDMA), and time division multiple access (TDMA)-based offloading schemes within UAV-enabled MEC systems, considering both infinite and finite blocklength scenarios. Through theoretically analysis of the minimum energy consumption required by these three schemes, two novel findings are presented. First, TDMA consistently achieves lower energy consumption than FDMA in both infinite and finite blocklength cases, due to the degrees of freedom afforded by sequential task offloading. Second, NOMA does not necessarily achieve lower energy consumption than FDMA when the offloading blocklength is finite, especially when the channel conditions and the offloaded task data sizes of two user equipments (UEs) are relatively symmetric. Furthermore, an alternating optimization algorithm that jointly optimizes the portions of task offloaded, the offloading times of all UEs, and the UAV location is proposed to solve the formulated energy consumption minimization problems. Simulation results verify the correctness of our analytical findings and demonstrate that the proposed algorithm effectively reduces MEC-related energy consumption compared to benchmark schemes that do not optimize task offloading portions and/or offloading times.
\end{abstract}


\section{Introduction}

\subsection{Background and Motivation}
Mobile edge computing (MEC) is regarded as a promising solution to resolving the conflict between resource-constrained user equipment (UE) and resource-hungry mobile applications by offloading computation tasks of UE to nearby computing servers at the network edge \cite{ref_MEC_Sabella}. In a conventional terrestrial MEC system, the task offloading performance is mainly limited by the poor wireless channels that typically suffer from various impairments such as shadowing and fading in addition to path loss. Moreover, in scenarios of limited or no communication infrastructure, such as disaster relief and emergency response \cite{ref_singleUAV_FDMA_Zhang}, a fixed terrestrial MEC server may not be able to sustain computation offloading for all UEs in a large area. Due to the advantages of unmanned aerial vehicles (UAVs) in terms of high mobility, easy deployment, line-of-sight (LoS) connections, and wide coverage \cite{ref_UAV,ref_UAV_Secure}, UAV-enabled MEC has been proposed as a potential technique to tackle the above challenges. Compared with its terrestrial counterpart, a UAV-carried MEC server can further enhance the flexibility and resilience of MEC and improve the UEs' experience \cite{ref_MEC_UAV}.

The energy efficiency is a key design aspect of a UAV-enabled MEC system, since both the UAV and the UEs are powered by batteries of a limited capacity \cite{ref_AG}. The MEC-related energy consumption includes the offloading energy consumption (i.e., communication energy consumption) of the UEs and the computation energy consumption of the UEs and the UAV \cite{ref_TE}. In a multi-UE UAV-enabled MEC system, both the communication and computation energy consumption will be affected by the multiple-access offloading scheduling and resource allocation \cite{ref_MA_MEC}. However, the energy efficiency of different multiple access offloading schemes has not been sufficiently studied for a UAV-enabled MEC system.

In addition, the fifth generation (5G) and beyond 5G (B5G) networks are expected to support enhanced mobile broadband (eMBB) and ultra-reliable low latency communication (URLLC) services \cite{ref_V1}. For eMBB services, the tasks offloaded by the UEs are packetized into long packets, e.g., thousands of bytes per packet, where the transmission blocklength can be assumed to be infinite and the decoding error can be assumed to be zero \cite{ref_IBL_FBL}. In this case, the Shannon capacity theorem is applicable to characterize the offloading rate. In contrast, for URLLC services, the size of a data packet is quite small, e.g., 20 or 32 bytes per packet, and the decoding errors in offloading transmission cannot be ignored due to the finite blocklength \cite{ref_URLLC}. Thus, the Shannon capacity cannot well characterize the offloading rate in the finite blocklength case \cite{ref_Poly}. Since the finite-blocklength transmission suffers from a lower data rate and a higher decoding error probability than the infinite-blocklength transmission, the multiple access and resource allocation schemes should be designed for these two transmission modes separately in a UAV-enabled MEC system, which remains a gap in the current literature. In particular, for non-orthogonal multiple access (NOMA), different from the infinite-blocklength transmission, the successive interference cancellation (SIC) failure probability should be considered in the finite-blocklength transmission \cite{ref_Sun,ref_Ren}, where the decoding performance of one UE's signal will be affected by the decoding error of the other UEs' signals that have been decoded earlier during SIC.

Motivated by the above, and to determine which multiple access offloading scheme achieves the highest energy efficiency in UAV-enabled MEC, this work aims to minimize and compare the MEC-related energy consumption. This is achieved by investigating various multiple access offloading schemes for both infinite and finite blocklength transmissions in a UAV-enabled MEC system.

\begin{table*}
	\begin{center}
		\caption{Comparison of this paper with related literature.}
		\label{tab1}
		\begin{tabular}{| c | c | c | c | c | c | c | c |}
			\hline
			\multirow{2}*{\bf{Reference}}& \multirow{2}*{\bf{UAV}}& \multicolumn{3}{|c|}{\bf{Multiple access scheme}}& \multicolumn{2}{|c|}{\bf{Blocklength}}& \multirow{2}*{\bf{Key findings}}\\
			\cline{3-7}
			{}& {}& NOMA& FDMA& TDMA& Infinite& Finite& {}\\
			\hline
			\multirow{2}*{\cite{ref_NOMA_Pan}}& {}& \multirow{2}*{$\checkmark$}& \multirow{2}*{$\checkmark$}& {}& \multirow{2}*{$\checkmark$}& {}& \parbox[t]{8.5cm}{NOMA achieves lower energy consumption than FDMA for tasks characterized by large data sizes and stringent delay requirements.}\\
			\hline 
			\multirow{2}*{\cite{ref_NOMA_Huang}}& {}& \multirow{2}*{$\checkmark$}& {}& \multirow{2}*{$\checkmark$}& \multirow{2}*{$\checkmark$}& {}& \parbox[t]{8.5cm}{Hybrid NOMA is more energy-efficient than TDMA for time-demanding tasks, but this performance advantage diminishes for delay-tolerant tasks.}\\
			\hline
			\multirow{2}*{\cite{ref_NOMA_Wang}}& {}& \multirow{2}*{$\checkmark$}& {}& \multirow{2}*{$\checkmark$}& \multirow{2}*{$\checkmark$}& {}& \parbox[t]{8.5cm}{NOMA achieves substantial energy consumption performance gains over TDMA-based offloading in multi-antenna MEC systems.}\\
			\hline
			\multirow{2}*{\cite{ref_NOMA_Yang}}& {}& \multirow{2}*{$\checkmark$}& \multirow{2}*{$\checkmark$}& \multirow{2}*{$\checkmark$}& \multirow{2}*{$\checkmark$}& {}& \parbox[t]{8.5cm}{When minimizing a linear combination of completion time and energy consumption, NOMA outperforms both TDMA and FDMA.}\\
			\hline
			\multirow{2}*{\cite{ref_Jeong}}& \multirow{2}*{$\checkmark$}& \multirow{2}*{$\checkmark$}& {}& \multirow{2}*{$\checkmark$}& \multirow{2}*{$\checkmark$}& {}& \parbox[t]{8.5cm}{In terms of total mobile energy consumption, NOMA offloading is preferable to TDMA, unless the computation delay requirements are stringent.}\\
			\hline
			\multirow{2}*{\cite{ref_NOMA_Zhang}}& \multirow{2}*{$\checkmark$}& \multirow{2}*{$\checkmark$}& \multirow{2}*{$\checkmark$}& {}& \multirow{2}*{$\checkmark$}& {}& \parbox[t]{8.5cm}{NOMA requires lower energy consumption than FDMA, and this advantage becomes more pronounced as the number of users increases.}\\
			\hline
			\multirow{3}*{\cite{ref_MEC_IRS}}& {}& \multirow{3}*{$\checkmark$}& {}& \multirow{3}*{$\checkmark$}& \multirow{3}*{$\checkmark$}& {}& \parbox[t]{8.5cm}{Due to the flexible IRS beamforming, TDMA can achieve a computation rate comparable to or even exceeding that of NOMA in an IRS-aided wireless-powered MEC system.}\\
			\hline
			\multirow{2}*{\cite{ref_IRS_Hu}}& {}& \multirow{2}*{$\checkmark$}& {}& {}& {}& \multirow{2}*{$\checkmark$}& \parbox[t]{8.5cm}{IRS-aided NOMA transmissions have been designed considering the practical implications of the finite-blocklength regime in MEC networks.}\\
			\hline
			\multirow{4}*{This paper}& \multirow{4}*{$\checkmark$}& \multirow{4}*{$\checkmark$}& \multirow{4}*{$\checkmark$}& \multirow{4}*{$\checkmark$}& \multirow{4}*{$\checkmark$}& \multirow{4}*{$\checkmark$}& \parbox[t]{8.5cm}{TDMA achieves lower energy consumption than FDMA. Furthermore, NOMA does not necessarily outperform FDMA when the offloading blocklength is finite, particularly when the channel conditions and offloaded task data sizes of the UEs are relatively symmetric.}\\
			\hline
		\end{tabular}
	\end{center}
\end{table*}

\subsection{Related Works}
Conventional orthogonal multiple access (OMA) schemes, such as frequency division multiple access (FDMA) and time division multiple access (TDMA), are commonly used in MEC systems to support multi-UE task offloading \cite{ref_singleUAV_FDMA_Zhang,ref_multiUAV_FDMA_Zhang,ref_MyFirst,ref_TDMA_Wang,ref_TDMA_OFDMA_You,ref_JiawenKang,ref_oneByone}. The average weighted energy consumption of the UEs and the UAV in a UAV-enabled MEC system was minimized in \cite{ref_singleUAV_FDMA_Zhang}, where all UEs were allocated the same bandwidth in FDMA-based task offloading. The computation efficiency of a multi-UAV-enabled MEC system was maximized in \cite{ref_multiUAV_FDMA_Zhang}, where each UAV serves its associated UEs with FDMA. In \cite{ref_MyFirst}, for a UAV-enabled MEC system with finite-blocklength FDMA offloading, the computation latency was minimized by jointly optimizing the computing times and central processing unit (CPU) frequencies of the UEs and the UAV, the offloading bandwidths of the UEs, and the three-dimensional location of the UAV. In the wireless-powered multi-UE MEC system proposed in \cite{ref_TDMA_Wang}, the UEs offloaded their tasks based on TDMA. The resource allocation strategies for both TDMA and orthogonal frequency-division multiple access (OFDMA) multi-UE MEC systems were studied in \cite{ref_TDMA_OFDMA_You} to maximize the computation capacity while reducing energy consumption and latency. In \cite{ref_JiawenKang}, the authors proposed a new solution that UAVs offload computation-intensive tasks to multiple edge servers with an effective multi-leader multi-follower Stackelberg game strategy. In \cite{ref_oneByone}, the UAV-aided MEC system allows the UAV to connect to only one UE at a time. In these OMA-based MEC systems, only a single UE is permitted to perform task offloading within each designated time/frequency resource block.

Due to the superposition coding at the transmitter and the SIC at the receiver, NOMA can achieve high spectral efficiency and massive connectivity for MEC \cite{ref_NOMA_MEC}. The resource allocation for latency minimization in NOMA-based MEC networks was investigated in \cite{ref_NOMA_Ding} and \cite{ref_NOMA_Fang}. In \cite{ref_NOMA_Ding}, the hybrid NOMA-MEC policy allows two UEs to offload their tasks in a shared time slot, wherein only one UE completes its task offloading while the other UE will offload its remaining task in the next time slot solely occupied by itself. While \cite{ref_NOMA_Fang} considered a pure NOMA offloading scheme, where multiple UEs can offload their tasks to the MEC server simultaneously in the same frequency band. The NOMA-based offloading was applied in \cite{ref_NOMA_Pan} and \cite{ref_NOMA_Huang} to minimize the energy consumption. Specifically, both the task uploading and result downloading in \cite{ref_NOMA_Pan} benefited from the advantages of NOMA, achieving energy-efficient MEC. In \cite{ref_NOMA_Huang}, the energy consumption of three MEC offloading strategies, including TDMA, pure NOMA, and hybrid NOMA, was investigated for a general and arbitrary network status. The multi-antenna NOMA technique was integrated into multi-UE computation offloading in \cite{ref_NOMA_Wang}, which jointly optimized the resource allocation and the decoding order for SIC to minimize the weighted sum energy consumption of all UEs. In \cite{ref_NOMA_Yang}, a linear combination of the computation time and total energy consumption was minimized for a user-group NOMA-MEC system. In \cite{ref_Jeong}, the total energy consumption of the UEs in a frequency-division-duplex UAV-enabled MEC system was minimized for both the TDMA and NOMA offloading schemes. To extend the MEC coverage and improve the system flexibility, NOMA was also considered in the multi-UAV-enabled MEC system \cite{ref_NOMA_Zhang}. The aforementioned studies demonstrated the performance advantages of NOMA over OMA in MEC systems. However, these investigations typically assumed an infinite offloading blocklength and perfect SIC for NOMA. Although \cite{ref_IRS_Hu} studied the energy efficiency of NOMA-based MEC in the finite blocklength regime, SIC was still assumed to be error free.

However, we note that NOMA does not always achieve performance advantages over OMA. For example, the throughput regions of OMA and NOMA are identical in the symmetric channel case \cite{ref_NOMA_chan}. In the intelligent reflecting surface (IRS)-assisted downlink communication network \cite{ref_Zheng}, TDMA is more energy-efficient than NOMA when two nearby users are paired and their rates are symmetric, since the IRS phase shift vector can be set differently for the two users in different time slots of TDMA. Furtherly, an IRS-aided wireless-powered MEC system was studied in \cite{ref_MEC_IRS}, and the authors unveiled that the TDMA offloading scheme can achieve a better computation rate than that of NOMA when the IRS beamforming is flexibly adapted for uplink offloading. Table~\ref{tab1} provides a comparative summary of system models and contributions, contrasting our work with existing state-of-the-art multiple-access MEC literature. From this comparison, it is evident that comprehensive studies on UAV-enabled MEC systems employing finite-blocklength NOMA offloading are still missing. On the other hand, it remains unknown yet whether NOMA can outperform OMA or not (e.g., in terms of energy consumption) in a UAV-enabled MEC system when the transmission blocklength is finite.

\subsection{Contributions}
In this paper, we investigate the MEC-related energy consumption minimization problem for a two-UE UAV-enabled MEC system while considering three different multiple access offloading schemes, i.e., NOMA, FDMA, and TDMA, for both the infinite and finite blocklength cases. Specifically, the MEC-related energy consumption is defined as the sum of the energy consumption caused by the local computing at the UEs, remote computing at the UAV, and the UEs' transmission for task offloading, and it is minimized by jointly optimizing the portions of task offloaded, the offloading times of all UEs, and the UAV location for each multiple access scheme. Our main contributions are summarized as follows:
\begin{itemize}
  \item{For each considered offloading scheme, the expression of the minimum required transmit power of each UE that can satisfy the information-causality constraint for computation offloading is derived, and the energy minimization problem that subjects to the computation latency constraint, maximum CPU frequency constraint, and maximum transmit power constraint is formulated. Notably, in the finite-blocklength NOMA offloading scheme, since the perfect SIC cannot be guaranteed, both SIC success and failure cases are considered, and a closed-form expression for the average minimum transmit power of these two cases is obtained.}
  \item{The relationships of minimum MEC-related energy consumption required by the considered multiple access offloading schemes are analyzed, and the analytical results show that TDMA can always achieve a lower energy consumption than FDMA due to the freedom of remote-computing time brought by the sequential task offloading of the UEs. In addition, NOMA can achieve a lower energy consumption than FDMA in the infinite blocklength case, but the same conclusion cannot be reached in the finite blocklength case since the SIC failure should be considered, especially when the channel conditions and the offloaded task data sizes of the two UEs are relatively symmetric.}
  \item{Since the formulated MEC-related energy consumption minimization problem is non-convex, it is solved by devising an alternating optimization algorithm, where the original problem is decomposed into three subproblems that optimize the portion of task offloaded by each UE, the offloading time of each UE, and the UAV location, respectively, based on the block coordinate descent (BCD) method. Furthermore, in the subproblems, the expressions of the minimum transmit power are transformed into a form containing the difference of two convex functions, which are handled by the successive convex approximation (SCA) technique.}
  \item{Simulation results verify the analytical results and show that the proposed algorithm can effectively reduce the MEC-related energy consumption, as compared with the benchmark schemes without optimizing the portions of task offloaded or the offloading times. The results show that the advantage of joint optimization is more markable when the computational burden is heavier or the computation latency requirement is more stringent for each UE. The results also show that the decoding error probability has a significant impact on NOMA offloading with finite-blocklength, due to nonnegligible SIC failures in this case.}
\end{itemize}

The rest of this paper is organized as follows. Section \uppercase\expandafter{\romannumeral2} presents the system models and problem formulations of all the considered multiple access offloading schemes in both infinite blocklength and finite blocklength cases. Section \uppercase\expandafter{\romannumeral3} provides the theoretical performance comparison of difference offloading schemes. Section \uppercase\expandafter{\romannumeral4} proposes the alternating optimization algorithm. Section \uppercase\expandafter{\romannumeral5} shows the simulation results to evaluate the effectiveness of the proposed algorithm. Section \uppercase\expandafter{\romannumeral6} concludes the paper.

\section{System Model and Problem Formulation}
\begin{figure}[!t]
  \centering
  \includegraphics[width=2.5in]{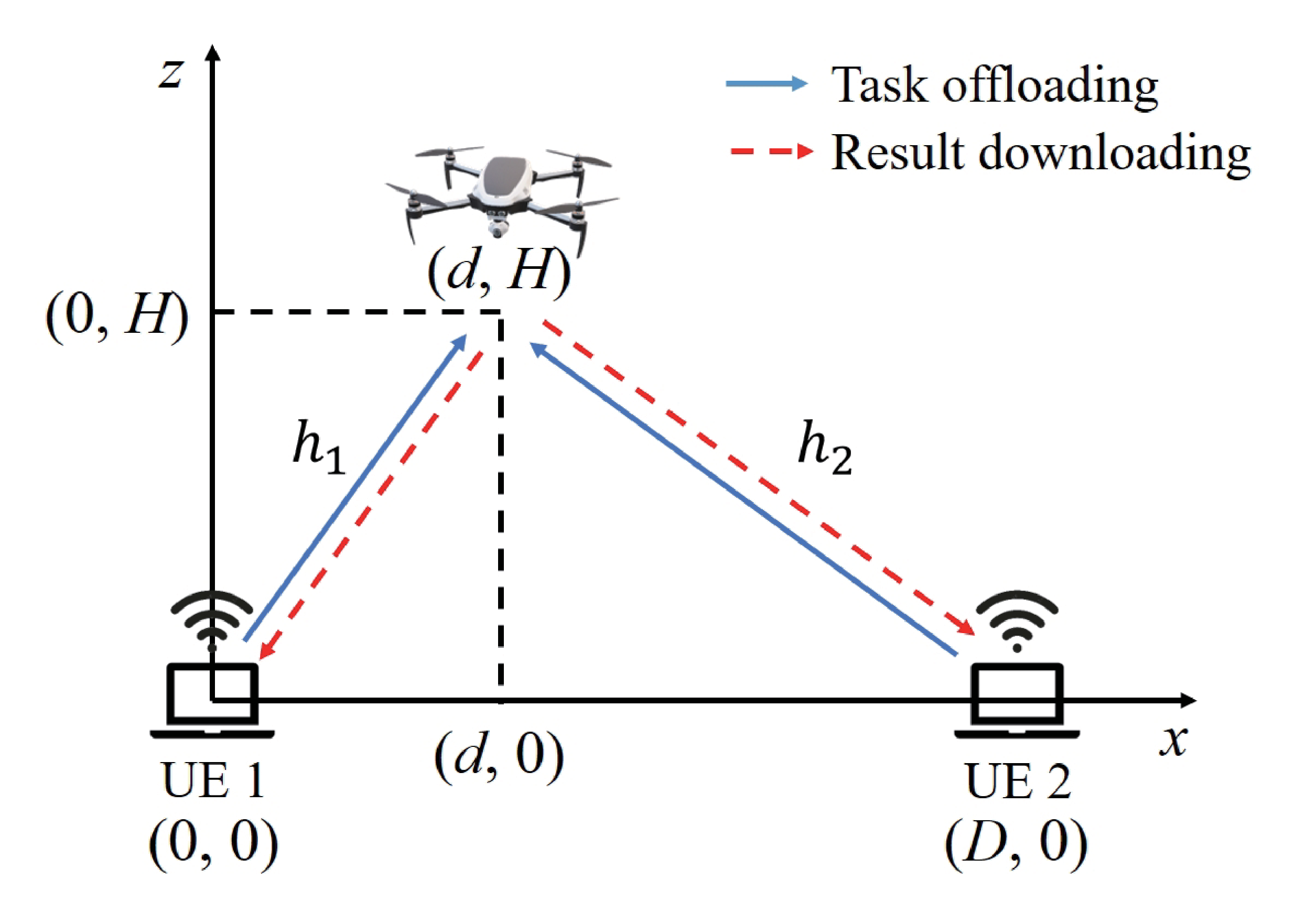}
  \caption{A two-UE UAV-enabled MEC system.}
  \label{fig_system}
\end{figure}
As shown in Fig. \ref{fig_system}, we consider a two-UE UAV-enabled MEC system, where the two UEs on the ground can offload parts of their computation tasks to a rotary-wing UAV that carries an MEC server for remote computing, and the UAV and UEs are each equipped with a single antenna. UE $k \in \{1,2\}$ is required to compute an $L_k$-bit task in a given time length $T_\mathrm{max}$. To minimize the sum transmission distance of the two offloading links, the UAV hovers above the horizontal line between the two UEs. Accordingly, the locations of the UAV and UEs are contained within a vertical two-dimensional Cartesian coordinate system. Without loss of generality, we consider that UE 1 is located at the origin $(0,0)$, and the locations of the UAV and UE 2 are $(d,H)$ and $(D,0)$, respectively, where $d$, $D$, and $H$ denote the horizontal distance between the UAV and UE 1, the distance between the two UEs, and the fixed altitude of the UAV, respectively.

It is worth pointing out that the considered two-UE system can be extended to scenarios of more than two UEs, where the UEs can be grouped into pairs and our proposed system can be applied for each pair. And if different UE pairs are allocated with different orthogonal time/frequency resources, they can offload their task to the UAV without multi-pair interference. Furthermore, by designing a two-timescale framework, we can also extend the considered system to a mobility scenario where the UAV provides computation offloading services for a wide region through cruising flight. Specifically, the UAV’s trajectory is optimized in the large timescale, while the computation offloading and resource allocation are optimized in the small timescale.

Since UAV typically has a high probability of establishing an LoS air-to-ground communication link \cite{ref_UAV}, the free space channel model can be adopted. Thus, the channel power gains from UEs 1 and 2 to the UAV can be represented as
\begin{equation}
  \label{deqn_ex1a}
   h_1 = \frac{\beta_0}{H^2 + d^2}, \; h_2 = \frac{\beta_0}{H^2 + (D-d)^2},
\end{equation}
respectively, where $\beta_0$ is the channel power gain at a reference distance of $d_0 = 1$~m. To design the communication and computation coordination between the UAV and UEs, a central controller can be deployed in the MEC network to collect the global channel state information (CSI) and computation-related information. Assuming a quasi-static channel environment for the duration of low-latency task offloading, perfect CSI is assumed to be known for the remainder of this paper to facilitate tractable analysis and derive fundamental performance benchmarks.\footnote{To improve the robustness and account for CSI error, a bounded error model can be considered to minimize the worst-case MEC-related energy consumption.}

The partial computation offloading mode is considered in the MEC. Denoting the portion of task offloaded by UE $k$ by $0 \leq \rho_k \leq 1$, the data sizes of local-computing and remote-computing can be expressed as $(1-\rho_k) L_k$ and $\rho_k L_k$, respectively. We consider scenarios such as factory automation and wireless federated learning, where the computation results are usually some control commands or model parameters whose data size is typically negligible compared to that of the offloaded task \cite{ref_Ren,ref_WFL}. For example, event-triggered control based on binary neural networks for image recognition outputs a binary parameter to indicate the presence of a target object~\cite{ref_BNN}, and the sparse gradient triggering mechanism in federated learning uploads a 1-bit signal to indicate the need for global model update~\cite{ref_SGD}. Thus, we do not consider the latency and energy consumption caused by the result downloading process from the UAV to UEs. Thus, if UE $k$'s task offloading time is $t_k$, the available execution times of its local and remote computing are $T_{\mathrm{max}}$ and $T_{\mathrm{max}} - t_k$, respectively.

By applying the dynamic voltage and frequency scaling (DVFS) technique, the UEs and UAV can control the energy consumption for local and remote computing by adjusting their CPU frequencies for each cycle, respectively~\cite{ref_DVFS}. To minimize the computation-related energy consumption, based on \cite{ref_TDMA_Wang}, the CPU frequencies for UE $k$'s local and remote computing during their available execution times should be adjusted as
\begin{equation}
  \label{deqn_ex2a}
  f_{k}^{\mathrm{loc}} = \frac{(1-\rho_k) c_k L_k}{T_{\mathrm{max}}}, \; f_{k}^{\mathrm{rem}} = \frac{\rho_k c_k L_k}{T_{\mathrm{max}} - t_k},
\end{equation}
respectively, where $c_k$ denotes the number of CPU cycles required to compute 1 bit of UE $k$'s task. Then, the energy consumption of UE $k$'s local computing and remote computing at the UAV can be characterized as
\begin{subequations}\label{deqn_ex3a}
  \begin{align}
   E_k^{\mathrm{loc}} &= \kappa_k T_{\mathrm{max}} (f_{k}^{\mathrm{loc}})^3, \label{deqn_ex3a_A}\\
   E_k^{\mathrm{rem}} &= \kappa_{\mathrm{U}} (T_{\mathrm{max}} - t_k) (f_{k}^{\mathrm{rem}})^3, \label{deqn_ex3a_B}
  \end{align}
\end{subequations}
respectively, where $\kappa_k$ and $\kappa_{\mathrm{U}}$ denote the effective capacitance coefficients of and UE $k$ and the UAV, respectively.

The energy consumption of each UE includes local-computing energy consumption and task-offloading energy consumption, and that of UAV mainly includes remote-computing energy consumption and its propulsion energy consumption. According to \cite{ref_UAVEnergy}, the propulsion energy consumption of the UAV is determined by physical parameters like aircraft weight, air density, rotor disc area, and so on and cannot be optimized by communication and computation resource allocation, so we only consider the communication-related and computation-related energy consumption and define the MEC-related energy consumption of UE $k$ as
\begin{equation}
  \label{deqn_ex4a}
  E_k^{\mathrm{MEC}} = E_k^{\mathrm{loc}} + E_k^{\mathrm{rem}} + E_k^{\mathrm{off}},
\end{equation}
where $E_k^{\mathrm{off}} = P_k t_k$ is the energy consumption of offloading, and $P_k$ is UE $k$'s transmit power.

It can be observed in \eqref{deqn_ex4a} that the MEC-related energy consumption of the UEs is determined by the portions of task offloaded, transmit powers, and offloading times of the UEs, which are related to the multiple access schemes in the task offloading. We consider three main multiple access schemes, namely NOMA, FDMA, and TDMA. These three multiple access offloading schemes will be sequentially modeled and analyzed below in both the infinite blocklength and finite blocklength cases, which correspond to the computation-intensive tasks in eMBB services and the mission-critical tasks in URLLC services, respectively. Additionally, the UAV location directly impacts the quality of the offloading channels, making it a crucial factor in designing a computation offloading strategy that balances communication-related energy consumption and computation-related energy consumption.

\begin{figure}[!t]
  \centering
  \includegraphics[width=3.5in]{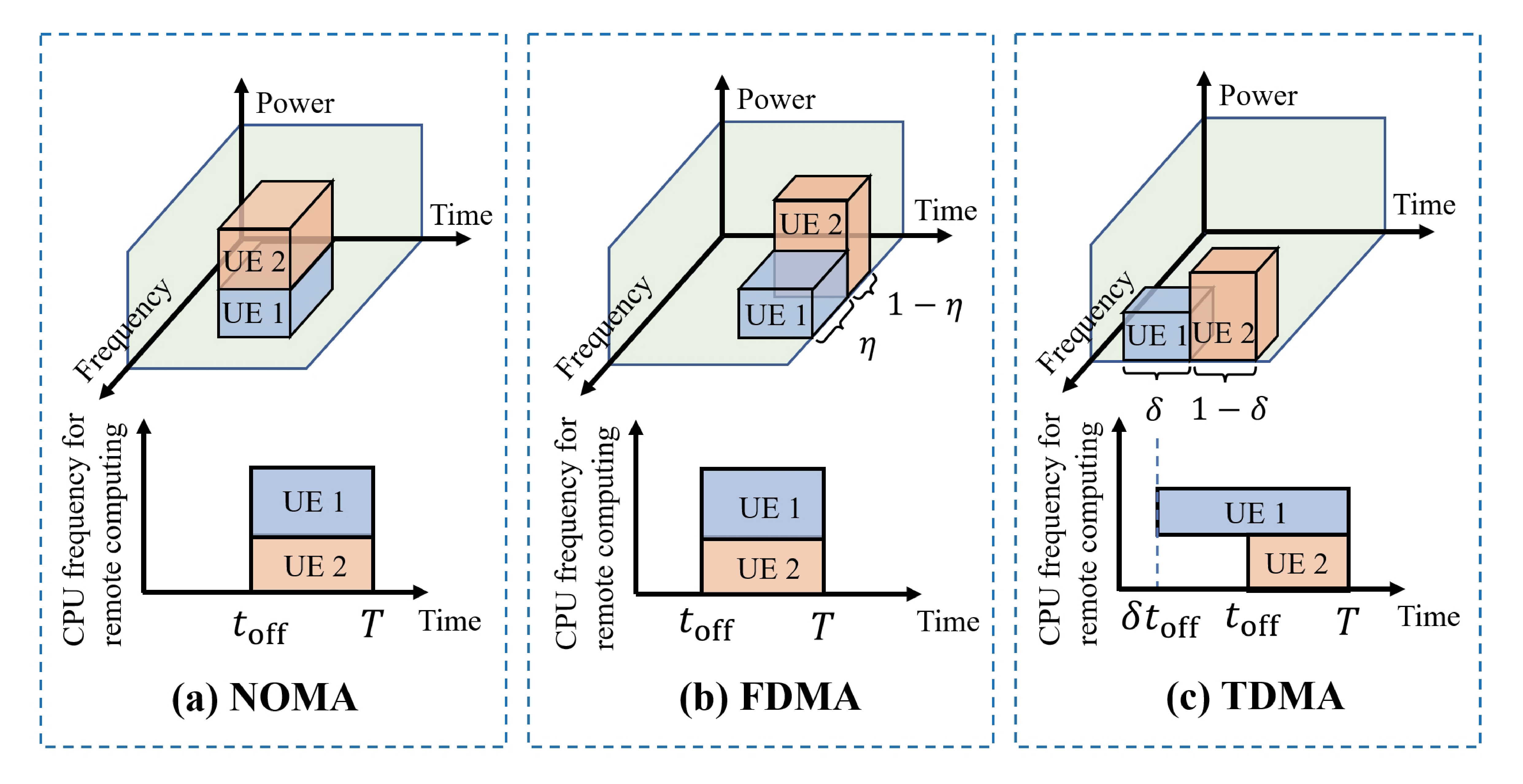}
  \caption{Illustrations of the time-frequency transmit power allocations and the remote computing CPU frequency allocations in different offloading schemes.}
  \label{fig_allocation}
\end{figure}

\subsection{NOMA Offloading}
By using NOMA, the UEs can offload tasks to the UAV simultaneously over the same time-frequency resource block as shown in Fig. \ref{fig_allocation}(a).\footnote{Power-domain NOMA is considered in this paper, where the UAV exploits differences in UEs' received power levels to distinguish between different UEs based on SIC.} As such, the offloading bandwidths and times of two UEs can be expressed as $B_1 = B_2 = B$ and $t_1 = t_2 = t$, respectively, where $B$ is the total bandwidth and $t$ is the offloading time. By letting $x_k \in \mathbb{C}$ denote the UE~$k$'s task-bearing signal with unit power for uplink offloading, i.e., $\mathbb{E}\left[|x_k|^2\right] = 1$, the received signal $y\in \mathbb{C}$ at the UAV is then expressed as
\begin{equation}
	\label{deqn_ex1z}
	y_{\mathrm{NOMA}} = \sqrt{P_1}h_1 x_1 + \sqrt{P_2}h_2 x_2 + z_{\mathrm{NOMA}},
\end{equation}
where $z_{\mathrm{NOMA}}\sim \mathcal{CN}(0,\sigma_{\mathrm{NOMA}}^2)$ denotes the additive white Gaussian noise (AWGN) at the UAV with $\sigma_{\mathrm{NOMA}}^2 = BN_0$ and $N_0$ being the noise power and the noise power spectrum density, respectively. The UAV performs SIC to decode two UEs' signals in descending order of their channel power gains. Note that since UEs 1 and 2's channel power gains $h_1$ and $h_2$ depend on the UAV location, we may have $2! = 2$ permutations for the decoding order, i.e., (i) the UAV first decodes UE 1's signal and then decodes UE 2's signal, and (ii) the UAV first decodes UE 2's signal and then decodes UE 1's signal. In the following, we choose order (i) to analyze, and the other order can be analyzed similarly.

\textit{1) Infinite Blocklength Case:} In this case, the transmission blocklength of two UEs is assumed to be infinite, i.e., $N \to \infty$. Based on Shannon's channel coding theorem, the decoding error probability can be made arbitrarily small if the channel codes are infinitely long. Thus, the decoding error probability is negligible during the SIC process in the infinite-blocklength NOMA offloading scheme~\cite{ref_NOMA_Mag,ref_NOMA_allo}. Specifically, the UAV first decodes the signal from the stronger UE (i.e., UE 1), treating the signal from the weaker UE (i.e., UE 2) as interference. Once UE 1's signal is successfully decoded, it is reconstructed and canceled from the composite signal. The UAV then proceeds to decode UE 2's signal without suffering from any interference. Thus, the signal-to-interference-plus-noise ratio (SINR) of UE 1's signal and the signal-to-noise ratio (SNR) of UE 2's signal at UAV are expressed as
\begin{equation}
  \label{deqn_ex1b}
  \gamma_{\mathrm{I},1}^{\mathrm{NOMA}} = \frac{P_1 h_1}{P_2 h_2 + BN_0}, \; \gamma_{\mathrm{I},2}^{\mathrm{NOMA}} = \frac{P_2 h_2}{BN_0},
\end{equation}
respectively.\footnote{Accounting for the imperfection in SIC resulting from practical factors such as channel estimation errors, $\gamma_{\mathrm{I},2}^{\mathrm{NOMA}}$ should be remodeled to include a residual interference parameter similar to \cite{ref_SIC_Gupta} and \cite{ref_SIC_Mouni}.}

For remote computing at the UAV, the information-causality constraint should be satisfied, i.e., the offloaded task data can only be computed at the UAV if it has already been previously received from the UEs. The information-causality constraint for UE~$k$'s computation offloading can be expressed as
\begin{equation}
  \label{deqn_ex2b}
  Bt \mathrm{log}_{2} \left( 1 + \gamma_{\mathrm{I},k}^{\mathrm{NOMA}} \right) \geq \rho_k L_k,\; \forall k,
\end{equation}
where the left-hand side (LHS) of \eqref{deqn_ex2b} represents the effective offloading throughput from UE $k$ to the UAV, in which the offloading data rate is measured by the Shannon capacity formula, while the right-hand side (RHS) denotes the data size of UE $k$'s remote-computing task.

To ensure that UEs' offloaded tasks can be efficiently received by the UAV for remote computing in the quantity required, according to \eqref{deqn_ex2b}, the minimum transmit powers of UEs 1 and 2 are respectively obtained as
\begin{subequations}\label{deqn_ex4b}
  \begin{align}
  P_{\mathrm{I},1}^{\mathrm{NOMA}} &= \frac{\Upsilon_1(\rho_1,t) (\Upsilon_2(\rho_2,t)+1)}{\bar{h}_1}, \label{deqn_ex4b_A}\\
  P_{\mathrm{I},2}^{\mathrm{NOMA}} &= \frac{\Upsilon_2(\rho_2,t)}{\bar{h}_2}, \label{deqn_ex4b_B}
  \end{align}
\end{subequations}
where $\Upsilon_k^{\mathrm{I}}(\rho_k,t) \triangleq \mathrm{exp}\left( \frac{\mathrm{ln}2 \rho_k L_k}{Bt} \right) - 1$ is the minimum SINR/SNR, and $\bar{h}_k = \frac{h_k}{BN_0}$ denotes the normalized channel power gain from UE $k$ to the UAV. Note that we do not constrain any power allocation order between the two UEs, since the power allocation in NOMA may not necessarily follow the reverse order of the UEs' channel gains \cite{ref_NOMA_power}. As shown in \eqref{deqn_ex4b}, the transmit power of the UE depends on various factors including the UAV location, offloading time, and offloading task data sizes.

Thus, according to \eqref{deqn_ex2a}--\eqref{deqn_ex4a} and \eqref{deqn_ex4b}, the MEC-related energy consumption of NOMA offloading in the infinite blocklength case is expressed as
\begin{align}
  E_{\mathrm{I}}^{\mathrm{NOMA}}(\bm{\rho},t,d) &= \sum_{k=1}^{2}{E_k^{\mathrm{MEC}}} = \sum_{k=1}^{2} \left( \kappa_{\mathrm{U}}\frac{(\rho_k c_k L_k)^3}{(T_{\mathrm{max}} - t)^2} \right. \nonumber\\
  &\quad \left.+ \kappa_k \frac{((1-\rho_k)c_k L_k)^3}{{T_{\mathrm{max}}}^2} + P_{\mathrm{I},k}^{\mathrm{NOMA}} t \right), \label{deqn_ex5b}
\end{align}
and the problem that jointly optimizes the UAV location $d$, the portion of task offloaded by each UE $\bm{\rho} \triangleq \{\rho_1,\rho_2\}$, and the offloading time $t$ to minimize $E_{\mathrm{I}}^{\mathrm{NOMA}}(\bm{\rho},t,d)$ can be formulated as{\footnote{Note that since the altitude of the UAV $H$ is assumed to be fixed, the location of the UAV can be expressed by its horizontal distance $d$ only.}}
\begin{subequations}\label{eq:1}
  \begin{alignat}{2}
    \text{(NOMA-I)} \quad \min_{\bm{\rho}, t,d} \quad & E_{\mathrm{I}}^{\mathrm{NOMA}}(\bm{\rho},t,d) & \label{eq:1A}\\
    \mbox{s.t.} \quad 
    & f_{1}^{\mathrm{rem}} + f_{2}^{\mathrm{rem}} \leq f_{\mathrm{U,max}}, & \label{eq:1B}\\
    & f_{k}^{\mathrm{loc}} \leq f_{k,\mathrm{max}}, \; \forall k, & \label{eq:1C}\\
    & P_{\mathrm{I},k}^{\mathrm{NOMA}} \leq P_{k,\mathrm{max}}, \; \forall k, & \label{eq:1D}\\
    & 0 \leq \rho_k \leq 1, \; \forall k, & \label{eq:1E}\\
    & 0 \leq t \leq T_{\mathrm{max}}, & \label{eq:1F}\\
    & d \leq \frac{1}{2} D, & \label{eq:1G}
  \end{alignat}
\end{subequations}
where $f_{\mathrm{U,max}}$ and $f_{k,\mathrm{max}}$ are the maximum available CPU frequencies of the UAV and UE $k$, respectively, and $P_{k,\mathrm{max}}$ is the maximum transmit power of UE $k$. \eqref{eq:1B} ensures that the sum of remote-computing CPU frequencies allocated to the UEs does not exceed the maximum available CPU frequencies of the UAV. \eqref{eq:1C} and \eqref{eq:1D} mean that UE $k$'s CPU frequency $f_{k}^{\mathrm{loc}}$ and transmit power $P_{\mathrm{I},k}^{\mathrm{NOMA}}$ are upper bounded by $f_{k,\mathrm{max}}$ and $P_{k,\mathrm{max}}$, respectively. \eqref{eq:1E} and \eqref{eq:1F} are the value ranges of $\{\rho_k\}$ and $t$, respectively, ensure the tasks required to be computed will be completed within the given time. \eqref{eq:1G} exists to satisfy the assumptions that $h_1 \geq h_2$ and the UAV first decodes UE 1's signal during SIC as we analyzed above.

\textit{2) Finite Blocklength Case:} In this case, the offloading transmission blocklength of two UEs is finite and can be given by $N_1 = N_2 = N = Bt$ \cite{ref_URLLC}. Since the blocklength is short and the decoding error probability cannot be ignored, the offloading data rate should be characterized by the data rate expression derived in \cite{ref_Poly}. For a given decoding error probability $\epsilon_k$, the information-causality constraint for computation offloading can be expressed as
\begin{align}
  \hspace{-0.05cm}(1 - \epsilon_k)N \hspace{-0.05cm}\left(\hspace{-0.05cm} \underbrace{\mathrm{log}_2(1 + \gamma_{\mathrm{F},k}^{\mathrm{NOMA}}) - \sqrt{\frac{V_k}{N}}\frac{Q^{-1}(\epsilon_k)}{\mathrm{ln}2}}_{\text{finite-blocklength}\;\text{offloading}\;\text{data}\;\text{rate}}\hspace{-0.1cm}\right)\hspace{-0.1cm} \geq\hspace{-0.05cm} \rho_k L_k, \; \forall k, \label{deqn_ex1c}
\end{align}
where the LHS of \eqref{deqn_ex1c} represents the effective offloading throughput from UE $k$ to the UAV, $V_k$ is the channel dispersion that can be approximated as 1 for convenience \cite{ref_V1}, and $Q^{-1} \left ( \cdot  \right )$ is the inverse function of $Q(x) \triangleq \int_{x}^\infty \frac{1}{\sqrt{2\pi} } \,e^{-\frac{t^{2} }{2} }  \,\mathrm{d}t$. 

Note that the perfect SIC cannot be guaranteed if the blocklength is finite \cite{ref_Sun}, so both SIC success and failure should be considered. The UAV first decodes UE 1's signal while the interference caused by UE 2 is treated as noise, so the SINR of UE 1's signal at the UAV is expressed as 
\begin{equation}
  \label{deqn_ex3c}
  \gamma_{\mathrm{F},1}^{\mathrm{NOMA}} = \frac{P_1 h_1}{P_2 h_2 + BN_0}.
\end{equation}
If SIC succeeds, UE 1's signal is correctly decoded and completely canceled, the probability of which is $1-\epsilon_1$, and the SNR of UE 2's signal is expressed as
\begin{equation}
  \label{deqn_ex4c}
  \hat{\gamma}_{\mathrm{F},2}^{\mathrm{NOMA}} = \frac{P_2 h_2}{BN_0}.
\end{equation}
Otherwise, if SIC fails, UE 1's signal is not decoded correctly, the probability of which is $\epsilon_1$, it cannot be successfully canceled, and thus the SINR of UE 2's signal is expressed as
\begin{equation}
  \label{deqn_ex5c}
  \check{\gamma}_{\mathrm{F},2}^{\mathrm{NOMA}} = \frac{P_2 h_2}{P_1 h_1 + BN_0}.
\end{equation}

By letting $\gamma_{\mathrm{F},2}^{\mathrm{NOMA}} = \hat{\gamma}_{\mathrm{F},2}^{\mathrm{NOMA}}$ and combining \eqref{deqn_ex1c}, \eqref{deqn_ex3c}, and \eqref{deqn_ex4c}, the minimum transmit powers of UEs 1 and 2 if the SIC succeeds can be obtained as
\begin{subequations}\label{deqn_ex6c}
  \begin{align}
  \hat{P}_{\mathrm{F},1}^{\mathrm{NOMA}} &= \frac{\Upsilon_1^{\mathrm{F}}(\rho_1,N) (\Upsilon_2^{\mathrm{F}}(\rho_2,N)+1)}{\bar{h}_1}, \label{deqn_ex6c_A}\\
  \hat{P}_{\mathrm{F},2}^{\mathrm{NOMA}} &= \frac{\Upsilon_2^{\mathrm{F}}(\rho_2,N)}{\bar{h}_2}, \label{deqn_ex6c_B}
  \end{align}
\end{subequations}
respectively, where $\Upsilon_k^{\mathrm{F}}(\rho_k,N) \triangleq \mathrm{exp} \left( \frac{\mathrm{ln}2 \rho_k L_k}{N(1 - \epsilon_k)} + \frac{Q^{-1}(\epsilon_k)}{\sqrt{N}} \right) - 1$ is the minimum SINR/SNR. If the SIC fails, by letting $\gamma_{\mathrm{F},2}^{\mathrm{NOMA}} = \check{\gamma}_{\mathrm{F},2}^{\mathrm{NOMA}}$ and combining \eqref{deqn_ex1c}, \eqref{deqn_ex3c}, and \eqref{deqn_ex5c}, the minimum transmit powers of UEs 1 and 2 can be respectively derived as
\begin{subequations}\label{deqn_ex7c}
  \begin{align}
   \check{P}_{\mathrm{F},1}^{\mathrm{NOMA}} &= \frac{1}{\bar{h}_1} \frac{\Upsilon_1^{\mathrm{F}}(\rho_1,N) (\Upsilon_2^{\mathrm{F}}(\rho_2,N)+1)}{1 - \Upsilon_1^{\mathrm{F}}(\rho_1,N)\Upsilon_2^{\mathrm{F}}(\rho_2,N)}, \label{deqn_ex7c_A}\\
   \check{P}_{\mathrm{F},2}^{\mathrm{NOMA}} &= \frac{1}{\bar{h}_2} \frac{\Upsilon_2^{\mathrm{F}}(\rho_2,N) (\Upsilon_1^{\mathrm{F}}(\rho_1,N)+1)}{1 - \Upsilon_1^{\mathrm{F}}(\rho_1,N)\Upsilon_2^{\mathrm{F}}(\rho_2,N)}. \label{deqn_ex7c_B}
  \end{align}
\end{subequations}
By combining the two cases, the average minimum transmit power of UE $k$ can be characterized as
\begin{equation}
  \label{deqn_ex8c}
  \bar{P}_{\mathrm{F},k}^{\mathrm{NOMA}} = (1 - \epsilon_1)\hat{P}_{\mathrm{F},k}^{\mathrm{NOMA}} + \epsilon_1 \check{P}_{\mathrm{F},k}^{\mathrm{NOMA}}.
\end{equation}

Thus, according to \eqref{deqn_ex2a}--\eqref{deqn_ex4a} and \eqref{deqn_ex8c}, the MEC-related energy consumption of NOMA offloading in the finite blocklength case is expressed as
\begin{align}
  E_{\mathrm{F}}^{\mathrm{NOMA}}(\bm{\rho},t,d) &= \sum_{k=1}^{2} \left( \kappa_{\mathrm{U}}\frac{(\rho_k c_k L_k)^3}{(T_{\mathrm{max}} - t)^2} \right. \nonumber\\
  &\left.+ \kappa_k \frac{((1-\rho_k)c_k L_k)^3}{{T_{\mathrm{max}}}^2} + \bar{P}_{\mathrm{F},k}^{\mathrm{NOMA}} t \right), \label{deqn_ex9c}
\end{align}
and the joint optimization problem that minimizing $E_{\mathrm{F}}^{\mathrm{NOMA}}(\bm{\rho},t,d)$ can be formulated as
\begin{subequations}\label{eq:2}
  \begin{alignat}{2}
    \text{(NOMA-F)} \quad \min_{\bm{\rho}, t,d} \quad & E_{\mathrm{F}}^{\mathrm{NOMA}}(\bm{\rho},t,d) & \label{eq:2A}\\
    \mbox{s.t.} \quad 
    & \bar{P}_{\mathrm{F},k}^{\mathrm{NOMA}} \leq P_{k,\mathrm{max}}, \; \forall k, & \label{eq:2B}\\
    & \Upsilon_1^{\mathrm{F}}(\rho_1,N)\Upsilon_2^{\mathrm{F}}(\rho_2,N) < 1, & \label{eq:2C}\\
    & \eqref{eq:1B}, \eqref{eq:1C}, \eqref{eq:1E}-\eqref{eq:1G}, & \label{eq:2D}
  \end{alignat}
\end{subequations}
where \eqref{eq:2C} ensures that $\check{P}_{\mathrm{F},k}^{\mathrm{NOMA}} \geq 0, \forall k$.

\subsection{FDMA Offloading}
By using FDMA, two UEs offload their tasks to the UAV simultaneously over two different frequency-domain resource blocks as shown in Fig. \ref{fig_allocation}(b). Thus, they have the same offloading time, i.e., $t_1 = t_2 = t$, and the bandwidths allocated to UEs 1 and 2 can be written as $B_1 = \eta B$ and $B_2 = (1-\eta)B$, respectively, where $0 \leq \eta \leq 1$ denotes the bandwidth allocation coefficient. The signal transmitted from UE~$k$ and received by the UAV is given by
\begin{equation}
	\label{deqn_ex2z}
	y_{k,\mathrm{FDMA}} = \sqrt{P_k}h_k x_k + z_{k,\mathrm{FDMA}},\;\forall k,
\end{equation}
where $z_{k,\mathrm{FDMA}}\sim \mathcal{CN}(0,\sigma_{k,\mathrm{FDMA}}^2)$ with $\sigma_{1,\mathrm{FDMA}}^2 = \eta BN_0$ and $\sigma_{2,\mathrm{FDMA}}^2 = (1-\eta)BN_0$. Accordingly, the SNRs of UEs 1 and 2's signals received by the UAV are expressed as
\begin{equation}
  \label{deqn_ex1d}
  \gamma_1^{\mathrm{FDMA}} = \frac{P_1 h_1}{\eta B N_0}, \; \gamma_2^{\mathrm{FDMA}} = \frac{P_2 h_2}{(1-\eta)B N_0},
\end{equation}
respectively.

\textit{1) Infinite Blocklength Case:} In this case, the information-causality constraint for computation offloading can be expressed as 
\begin{equation}
  \label{deqn_ex2d}
  B_k t \mathrm{log}_{2} \left( 1 + \gamma_{\mathrm{I},k}^{\mathrm{FDMA}} \right) \geq \rho_k L_k,\; \forall k,
\end{equation}
where $\gamma_{\mathrm{I},k}^{\mathrm{FDMA}} = \gamma_{k}^{\mathrm{FDMA}}$. According to \eqref{deqn_ex2d}, the minimum transmit powers of UEs 1 and 2 are respectively obtained as
\begin{subequations}\label{deqn_ex4d}
  \begin{align}
  P_{\mathrm{I},1}^{\mathrm{FDMA}} &= \frac{\eta}{\bar{h}_1}\Upsilon_1^{\mathrm{I}}(\rho_1,\eta t), \label{deqn_ex4d_A}\\
  P_{\mathrm{I},2}^{\mathrm{FDMA}} &= \frac{1-\eta}{\bar{h}_2}\Upsilon_2^{\mathrm{I}}(\rho_2,(1-\eta) t). \label{deqn_ex4d_B}
  \end{align}
\end{subequations}

Thus, according to \eqref{deqn_ex2a}--\eqref{deqn_ex4a} and \eqref{deqn_ex4d}, the MEC-related energy consumption of FDMA offloading in the infinite blocklength case is expressed as
\begin{align}
  E_{\mathrm{I}}^{\mathrm{FDMA}}(\bm{\rho},t,d) &= \sum_{k=1}^{2} \left( \kappa_{\mathrm{U}}\frac{(\rho_k c_k L_k)^3}{(T_{\mathrm{max}} - t)^2} \right. \nonumber\\
  &\left.+ \kappa_k \frac{((1-\rho_k)c_k L_k)^3}{{T_{\mathrm{max}}}^2} + P_{\mathrm{I},k}^{\mathrm{FDMA}} t \right), \label{deqn_ex5d}
\end{align}
and the joint optimization problem that minimizing $E_{\mathrm{I}}^{\mathrm{FDMA}}(\bm{\rho},t,d)$ can be formulated as
\begin{subequations}\label{eq:3}
  \begin{alignat}{2}
    \text{(FDMA-I)} \quad \min_{\bm{\rho}, t,d} \quad & E_{\mathrm{I}}^{\mathrm{FDMA}}(\bm{\rho},t,d) & \label{eq:3A}\\
    \mbox{s.t.} \quad 
    & P_{\mathrm{I},k}^{\mathrm{FDMA}} \leq P_{k,\mathrm{max}}, \; \forall k, & \label{eq:3B}\\
    & \eqref{eq:1B}, \eqref{eq:1C}, \eqref{eq:1E},\eqref{eq:1F}. & \label{eq:3C}
  \end{alignat}
\end{subequations}

\textit{2) Finite Blocklength Case:} In this case, the blocklength of UEs 1 and 2 can be written as $N_1 = B_1 t = \eta N$ and $N_2 = B_2 t = (1-\eta) N$, respectively. And for a given decoding error probability $\epsilon_k$, the information-causality constraint for computation offloading can be expressed as
\begin{align}
  (1 - \epsilon_k)N_k &\left( \mathrm{log}_2(1 + \gamma_{\mathrm{F},k}^{\mathrm{FDMA}}) - \sqrt{\frac{V_k}{N_k}}\frac{Q^{-1}(\epsilon_k)}{\mathrm{ln}2}\right) \geq \nonumber\\
  &\rho_k L_k, \; \forall k, \label{deqn_ex1e}
\end{align}
where $\gamma_{\mathrm{F},k}^{\mathrm{FDMA}} = \gamma_{k}^{\mathrm{FDMA}}$. According to \eqref{deqn_ex1e}, the minimum transmit powers of UEs 1 and 2 are respectively obtained as
\begin{subequations}\label{deqn_ex3e}
  \begin{align}
  P_{\mathrm{F},1}^{\mathrm{FDMA}} &= \frac{\eta}{\bar{h}_1}\Upsilon_1^{\mathrm{F}}(\rho_1,\eta N), \label{deqn_ex3e_A}\\
  P_{\mathrm{F},2}^{\mathrm{FDMA}} &= \frac{1-\eta}{\bar{h}_2}\Upsilon_2^{\mathrm{F}}(\rho_2,(1-\eta) N). \label{deqn_ex3e_B}
  \end{align}
\end{subequations}

Thus, according to \eqref{deqn_ex2a}--\eqref{deqn_ex4a} and \eqref{deqn_ex3e}, the MEC-related energy consumption of FDMA offloading in the finite blocklength case is expressed as
\begin{align}
  E_{\mathrm{F}}^{\mathrm{FDMA}}(\bm{\rho},t,d) &= \sum_{k=1}^{2} \left( \kappa_{\mathrm{U}}\frac{(\rho_k c_k L_k)^3}{(T_{\mathrm{max}} - t)^2} \right. \nonumber\\
  &\left.+ \kappa_k \frac{((1-\rho_k)c_k L_k)^3}{{T_{\mathrm{max}}}^2} + P_{\mathrm{F},k}^{\mathrm{FDMA}} t \right), \label{deqn_ex4e}
\end{align}
and the joint optimization problem that minimizing $E_{\mathrm{F}}^{\mathrm{FDMA}}(\bm{\rho},t,d)$ can be formulated as
\begin{subequations}\label{eq:4}
  \begin{alignat}{2}
    \text{(FDMA-F)} \quad \min_{\bm{\rho}, t,d} \quad & E_{\mathrm{F}}^{\mathrm{FDMA}}(\bm{\rho},t,d) & \label{eq:4A}\\
    \mbox{s.t.} \quad 
    & P_{\mathrm{F},k}^{\mathrm{FDMA}} \leq P_{k,\mathrm{max}}, \; \forall k, & \label{eq:4B}\\
    & \eqref{eq:1B}, \eqref{eq:1C}, \eqref{eq:1E},\eqref{eq:1F}. & \label{eq:4C}
  \end{alignat}
\end{subequations}

\subsection{TDMA Offloading}
By using TDMA, as shown in Fig. \ref{fig_allocation}(c), two UEs offload their tasks to the UAV over two successive time-domain resource blocks, sharing the same bandwidth. Thus, the bandwidths of UEs 1 and 2 can be written as $B_1 = B_2 = B$, and their offloading times can be written as $t_1 = \delta t$ and $t_2 = (1-\delta)t$, respectively, where $0 \leq \delta \leq 1$ denotes the time allocation coefficient. The signal transmitted from UE~$k$ and received by the UAV is given by
	\begin{equation}
		\label{deqn_ex3z}
		y_{k,\mathrm{TDMA}} = \sqrt{P_k}h_k x_k + z_{\mathrm{TDMA}},\;\forall k,
	\end{equation}
	where $z_{\mathrm{TDMA}}\sim \mathcal{CN}(0,\sigma_{\mathrm{TDMA}}^2)$ with $\sigma_{\mathrm{TDMA}}^2 = BN_0$. Accordingly, the SNR of UE $k$'s signal received by the UAV is
\begin{equation}
  \label{deqn_ex1f}
  \gamma_{k}^{\mathrm{TDMA}} = \frac{P_k h_k}{B N_0}, \; \forall k.
\end{equation}

Note that it may have $2! = 2$ permutations for the offloading order, i.e., (i) UE 1 offloads its task before UE 2 and (ii) UE 2 offloads its task before UE 1. In the following, we choose order (i) to analyze, and the other order can be analyzed similarly. Different from the NOMA and FDMA offloadings, where the UAV computes two UEs' offloaded tasks simultaneously, the UAV can compute UE 1's offloaded task during UE 2 offloads its task in TDMA offloading. Thus, UEs 1 and 2's remote-computing times can be written as $T_{\mathrm{max}} - \delta t$ and $T_{\mathrm{max}} - t$, respectively, and the CPU frequencies allocated for their remote computing by the UAV can be expressed as
\begin{equation}
  \label{deqn_ex2f}
  f_{\mathrm{TDMA},1}^{\mathrm{rem}} = \frac{\rho_1 c_1 L_1}{T_{\mathrm{max}} - \delta t}, \; f_{\mathrm{TDMA},2}^{\mathrm{rem}} = \frac{\rho_2 c_2 L_2}{T_{\mathrm{max}} - t}.
\end{equation}

\textit{1) Infinite Blocklength Case:} In this case, the information-causality constraint for computation offloading can be expressed as 
\begin{equation}
  \label{deqn_ex3f}
  B t_k \mathrm{log}_{2} \left( 1 + \gamma_{\mathrm{I},k}^{\mathrm{TDMA}} \right) \geq \rho_k L_k\; \forall k,
\end{equation}
where $\gamma_{\mathrm{I},k}^{\mathrm{TDMA}} = \gamma_{k}^{\mathrm{TDMA}}$. According to \eqref{deqn_ex3f}, the minimum transmit powers of UEs 1 and 2 are respectively obtained as
\begin{subequations}\label{deqn_ex5f}
  \begin{align}
  P_{\mathrm{I},1}^{\mathrm{TDMA}} &= \frac{1}{\bar{h}_1}\Upsilon_1^{\mathrm{I}}(\rho_1,\delta t), \label{deqn_ex5f_A}\\
  P_{\mathrm{I},2}^{\mathrm{TDMA}} &= \frac{1}{\bar{h}_2}\Upsilon_2^{\mathrm{I}}(\rho_2,(1-\delta) t). \label{deqn_ex5f_B}
  \end{align}
\end{subequations}

Thus, according to \eqref{deqn_ex3a}, \eqref{deqn_ex4a}, \eqref{deqn_ex2f}, and \eqref{deqn_ex5f}, the MEC-related energy consumption of TDMA offloading in the infinite blocklength case is expressed as
\begin{align}
  E_{\mathrm{I}}^{\mathrm{TDMA}}(\bm{\rho},t,d) &= \kappa_{\mathrm{U}}\frac{(\rho_1 c_1 L_1)^3}{(T_{\mathrm{max}} - \delta t)^2} + \kappa_{\mathrm{U}}\frac{(\rho_2 c_2 L_2)^3}{(T_{\mathrm{max}} - t)^2} \nonumber\\
  & \quad + \delta P_{\mathrm{I},1}^{\mathrm{TDMA}} t + (1-\delta)P_{\mathrm{I},2}^{\mathrm{TDMA}} t \nonumber\\
  & \quad + \sum_{k=1}^{2}{\kappa_k \frac{((1-\rho_k)c_k L_k)^3}{{T_{\mathrm{max}}}^2}}, \label{deqn_ex6f}
\end{align}
and the joint optimization problem that minimizing $E_{\mathrm{I}}^{\mathrm{TDMA}}(\bm{\rho},t,d)$ can be formulated as
\begin{subequations}\label{eq:5}
  \begin{alignat}{2}
    \text{(TDMA-I)} \quad \min_{\bm{\rho}, t,d} \quad & E_{\mathrm{I}}^{\mathrm{TDMA}}(\bm{\rho},t,d) & \label{eq:5A}\\
    \mbox{s.t.} \quad 
    & f_{\mathrm{TDMA},1}^{\mathrm{rem}} + f_{\mathrm{TDMA},2}^{\mathrm{rem}} \leq f_{\mathrm{U,max}}, & \label{eq:5B}\\
    & P_{\mathrm{I},k}^{\mathrm{TDMA}} \leq P_{k,\mathrm{max}}, \; \forall k, & \label{eq:5C}\\
    & \eqref{eq:1C}, \eqref{eq:1E},\eqref{eq:1F}. & \label{eq:5D}
  \end{alignat}
\end{subequations}

\textit{2) Finite Blocklength Case:} In this case, the blocklength of UEs 1 and 2 can be written as $N_1 = Bt_1 = \delta N$ and $N_2 = Bt_2 = (1-\delta) N$, respectively. And for a given decoding error probability $\epsilon_k$, the information-causality constraint for computation offloading can be expressed as
\begin{align}
  (1 - \epsilon_k)N_k &\left( \mathrm{log}_2(1 + \gamma_{\mathrm{F},k}^{\mathrm{TDMA}}) - \sqrt{\frac{V_k}{N_k}}\frac{Q^{-1}(\epsilon_k)}{\mathrm{ln}2}\right) \geq \nonumber\\
  &\rho_k L_k, \; \forall k, \label{deqn_ex1g}
\end{align}
where $\gamma_{\mathrm{F},k}^{\mathrm{TDMA}} = \gamma_{k}^{\mathrm{TDMA}}$. According to \eqref{deqn_ex1g}, the minimum transmit powers of UEs 1 and 2 are respectively obtained as
\begin{subequations}\label{deqn_ex3g}
  \begin{align}
  P_{\mathrm{F},1}^{\mathrm{TDMA}} &= \frac{1}{\bar{h}_1}\Upsilon_1^{\mathrm{F}}(\rho_1,\delta N), \label{deqn_ex3g_A}\\
  P_{\mathrm{F},2}^{\mathrm{TDMA}} &= \frac{1}{\bar{h}_2}\Upsilon_2^{\mathrm{F}}(\rho_2,(1-\delta) N). \label{deqn_ex3g_B}
  \end{align}
\end{subequations}

Thus, according to \eqref{deqn_ex3a}, \eqref{deqn_ex4a}, \eqref{deqn_ex2f}, and \eqref{deqn_ex3g}, the MEC-related energy consumption of TDMA offloading in the finite blocklength case is expressed as
\begin{align}
  E_{\mathrm{F}}^{\mathrm{TDMA}}(\bm{\rho},t,d) &= \kappa_{\mathrm{U}}\frac{(\rho_1 c_1 L_1)^3}{(T_{\mathrm{max}} - \delta t)^2} + \kappa_{\mathrm{U}}\frac{(\rho_2 c_2 L_2)^3}{(T_{\mathrm{max}} - t)^2} \nonumber\\
  & \quad + \delta P_{\mathrm{F},1}^{\mathrm{TDMA}} t + (1-\delta)P_{\mathrm{F},2}^{\mathrm{TDMA}} t \nonumber\\
  & \quad + \sum_{k=1}^{2}{\kappa_k \frac{((1-\rho_k)c_k L_k)^3}{{T_{\mathrm{max}}}^2}}, \label{deqn_ex4g}
\end{align}
and the joint optimization problem that minimizing $E_{\mathrm{F}}^{\mathrm{TDMA}}(\bm{\rho},t,d)$ can be formulated as
\begin{subequations}\label{eq:6}
  \begin{alignat}{2}
    \text{(TDMA-F)} \quad \min_{\bm{\rho}, t,d} \quad & E_{\mathrm{F}}^{\mathrm{TDMA}}(\bm{\rho},t,d) & \label{eq:6A}\\
    \mbox{s.t.} \quad 
    & P_{\mathrm{F},k}^{\mathrm{TDMA}} \leq P_{k,\mathrm{max}}, \; \forall k, & \label{eq:6B}\\
    & \eqref{eq:1C}, \eqref{eq:1E}, \eqref{eq:1F}, \eqref{eq:5B}. & \label{eq:6C}
  \end{alignat}
\end{subequations}

\section{Performance Comparison}
To explore which multiple access offloading schemes is more capable of reducing the MEC-related energy consumption of the considered UAV-enabled MEC system and whether the differences in performance benefit among different schemes are the same in both infinite blocklength and finite blocklength cases, we compare the minimum MEC-related energy consumption obtained by NOMA, FDMA, and TDMA offloadings, respectively, in both infinite blocklength and finite blocklength cases. The analysis results of this section remain applicable in a multi-UE system, where UEs are grouped into pairs, and each UE pair is allocated with orthogonal time/frequency communication resources and independent CPU frequency for remote computing.

\subsection{Comparison between FDMA and TDMA Offloadings}
First, we compare the MEC-related energy consumption of FDMA and TDMA offloadings with finite blocklength. For the special case that $\eta = \delta$, let $\{\bm{\rho}_{\mathrm{F}}^{\ast}, t_{\mathrm{F}}^{\ast},d_{\mathrm{F}}^{\ast}\}$ denote the optimal solution to problem (FDMA-F) in \eqref{eq:4}, the energy gap between FDMA and TDMA, i.e., the difference between \eqref{deqn_ex4e} and \eqref{deqn_ex4g}, is given by 
\begin{align}
  \label{deqn_ex1h}
  & E^{\mathrm{FDMA}}_{\mathrm{F}}(\bm{\rho},t,d) - E^{\mathrm{TDMA}}_{\mathrm{F}}(\bm{\rho},t,d) \nonumber\\
  \overset{\text{(a)}}{\geq} & E^{\mathrm{FDMA}}_{\mathrm{F}}(\bm{\rho}_{\mathrm{F}}^{\ast}, t_{\mathrm{F}}^{\ast},d_{\mathrm{F}}^{\ast}) - E^{\mathrm{TDMA}}_{\mathrm{F}}(\bm{\rho}_{\mathrm{F}}^{\ast}, t_{\mathrm{F}}^{\ast},d_{\mathrm{F}}^{\ast}) \nonumber\\
  = & \kappa_{\mathrm{U}} \left(\frac{(\rho_{\mathrm{F},1}^{\ast} c_1 L_1)^3}{(T_{\mathrm{max}} - t_{\mathrm{F}}^{\ast})^2} - \frac{(\rho_{\mathrm{F},1}^{\ast} c_1 L_1)^3}{(T_{\mathrm{max}} - \delta t_{\mathrm{F}}^{\ast})^2} \right) \overset{\text{(b)}}{\geq} 0,
\end{align}
where the equality in (a) holds if $\{\bm{\rho}_{\mathrm{F}}^{\ast}, t_{\mathrm{F}}^{\ast},d_{\mathrm{F}}^{\ast}\}$ is also the optimal solution to problem (TDMA-F), and the equality in (b) holds if $\delta = 1$. Similarly, a non-negative energy gap between FDMA and TDMA offloadings with infinite blocklength can be also obtained. For the equality relation in \eqref{deqn_ex1h}, we find that FDMA and TDMA offloadings have the same task-offloading energy consumption $\{E_k^{\mathrm{off}}\}$ with any given $\{\bm{\rho},t,d\}$ and $\eta = \delta$. Furthermore, for the inequality relationship in \eqref{deqn_ex1h}, we find that TDMA can achieve a lower MEC-related energy consumption than FDMA due to the freedom of remote-computing time brought by the sequential task offloading of the UEs.

\subsection{Comparison between NOMA and FDMA Offloadings with Infinite Blocklength}
Next, we compare the MEC-related energy consumption of NOMA and FDMA offloadings with infinite blocklength. For a special case that $h_1 \geq h_2$ (the UAV first decodes UE 1's signal) and $\eta = \frac{1}{2}$, let $\{\tilde{\bm{\rho}}_{\mathrm{F}}^{\ast}, \tilde{t}_{\mathrm{F}}^{\ast},\tilde{d}_{\mathrm{F}}^{\ast}\}$ denote the optimal solution to the problem (FDMA-I) in \eqref{eq:3}, and $\{\tilde{h}_{\mathrm{F},k}^{\ast}\}$ denotes the corresponding normalized channel power gains, the energy gap between NOMA and FDMA, i.e., the difference between \eqref{deqn_ex5b} and \eqref{deqn_ex5d}, can be expressed as
\begin{align}
  & E^{\mathrm{NOMA}}_{\mathrm{I}}(\bm{\rho},t,d) - E^{\mathrm{FDMA}}_{\mathrm{I}}(\bm{\rho},t,d) \nonumber\\
  \overset{\text{(a)}}{\leq} & E^{\mathrm{NOMA}}_{\mathrm{I}}(\tilde{\bm{\rho}}_{\mathrm{F}}^{\ast}, \tilde{t}_{\mathrm{F}}^{\ast},\tilde{d}_{\mathrm{F}}^{\ast}) - E^{\mathrm{FDMA}}_{\mathrm{I}}(\tilde{\bm{\rho}}_{\mathrm{F}}^{\ast}, \tilde{t}_{\mathrm{F}}^{\ast},\tilde{d}_{\mathrm{F}}^{\ast}) \nonumber\\
  = & \left[ \frac{1}{\tilde{h}_{\mathrm{F},1}^{\ast}} \bar{\Upsilon}_1^{\mathrm{I}}(\tilde{\rho}_{\mathrm{F},1}^{\ast},\tilde{t}_{\mathrm{F}}^{\ast}) \bar{\Upsilon}_2^{\mathrm{I}}(\tilde{\rho}_{\mathrm{F},2}^{\ast},\tilde{t}_{\mathrm{F}}^{\ast}) + \left( \frac{1}{\tilde{h}_{\mathrm{F},2}^{\ast}} - \frac{1}{\tilde{h}_{\mathrm{F},1}^{\ast}}\right) \right. \nonumber\\
  & \quad \times \left(\bar{\Upsilon}_2^{\mathrm{I}}(\tilde{\rho}_{\mathrm{F},2}^{\ast},\tilde{t}_{\mathrm{F}}^{\ast}) - \frac{1}{2} \bar{\Upsilon}_2^{\mathrm{I}}(\tilde{\rho}_{\mathrm{F},2}^{\ast},\frac{\tilde{t}_{\mathrm{F}}^{\ast}}{2}) \right) \nonumber\\
  & \quad \left. - \frac{1}{2\tilde{h}_{\mathrm{F},1}^{\ast}}\sum_{k=1}^{2}{\bar{\Upsilon}_k^{\mathrm{I}}(\tilde{\rho}_{\mathrm{F},k}^{\ast},\frac{\tilde{t}_{\mathrm{F}}^{\ast}}{2})} + \left( \frac{1}{2\tilde{h}_{\mathrm{F},1}^{\ast}} - \frac{1}{2\tilde{h}_{\mathrm{F},2}^{\ast}}\right) \right] \tilde{t}_{\mathrm{F}}^{\ast} \nonumber\\
  \overset{\text{(b)}}{\leq} & \left[ \bar{\Upsilon}_1^{\mathrm{I}}(\tilde{\rho}_{\mathrm{F},1}^{\ast},\tilde{t}_{\mathrm{F}}^{\ast}) \bar{\Upsilon}_2^{\mathrm{I}}(\tilde{\rho}_{\mathrm{F},2}^{\ast},\tilde{t}_{\mathrm{F}}^{\ast}) - \sum_{k=1}^{2}{\frac{\bar{\Upsilon}_k^{\mathrm{I}}(\tilde{\rho}_{\mathrm{F},k}^{\ast},\frac{\tilde{t}_{\mathrm{F}}^{\ast}}{2})}{2}} \right] \frac{\tilde{t}_{\mathrm{F}}^{\ast}}{\tilde{h}_{\mathrm{F},1}^{\ast}} \nonumber\\
  = & \left[ \sqrt{\bar{\Upsilon}_1^{\mathrm{I}}(\tilde{\rho}_{\mathrm{F},1}^{\ast},\frac{\tilde{t}_{\mathrm{F}}^{\ast}}{2}) \bar{\Upsilon}_2^{\mathrm{I}}(\tilde{\rho}_{\mathrm{F},2}^{\ast},\frac{\tilde{t}_{\mathrm{F}}^{\ast}}{2})} - \sum_{k=1}^{2}{\frac{\bar{\Upsilon}_k^{\mathrm{I}}(\tilde{\rho}_{\mathrm{F},k}^{\ast},\frac{\tilde{t}_{\mathrm{F}}^{\ast}}{2})}{2}}  \right] \nonumber\\
  & \quad \times \frac{\tilde{t}_{\mathrm{F}}^{\ast}}{\tilde{h}_{\mathrm{F},1}^{\ast}} \overset{\text{(c)}}{\leq} 0, \label{deqn_ex2h}
\end{align}
where $\bar{\Upsilon}_k^{\mathrm{I}}(\rho_k,t) \triangleq \Upsilon_k^{\mathrm{I}}(\rho_k,t) + 1 = \mathrm{exp} ( \frac{\mathrm{ln}2 \rho_k L_k}{Bt})$. In \eqref{deqn_ex2h}, the inequality (b) holds due to $\tilde{h}_{\mathrm{F},1}^{\ast} \geq \tilde{h}_{\mathrm{F},2}^{\ast}$ and $\tilde{t}_{\mathrm{F}}^{\ast} \bar{\Upsilon}_2^{\mathrm{I}}(\tilde{\rho}_{\mathrm{F},2}^{\ast},\tilde{t}_{\mathrm{F}}^{\ast}) < \frac{\tilde{t}_{\mathrm{F}}^{\ast}}{2} \bar{\Upsilon}_2^{\mathrm{I}}(\tilde{\rho}_{\mathrm{F},2}^{\ast},\frac{\tilde{t}_{\mathrm{F}}^{\ast}}{2})$, where the latter can be proved by the fact that $f(x) = x e^{-x}$ is a decreasing function when $x \geq 1$, and then $B\left( \tilde{t}_{\mathrm{F}}^{\ast} \bar{\Upsilon}_2^{\mathrm{I}}(\tilde{\rho}_{\mathrm{F},2}^{\ast},\tilde{t}_{\mathrm{F}}^{\ast}) - \frac{\tilde{t}_{\mathrm{F}}^{\ast}}{2} \bar{\Upsilon}_2^{\mathrm{I}}(\tilde{\rho}_{\mathrm{F},2}^{\ast},\frac{\tilde{t}_{\mathrm{F}}^{\ast}}{2}) \right) = e^{\mathrm{ln}2 \tilde{\rho}_{\mathrm{F},2}^{\ast} L_2} \left( B\tilde{t}_{\mathrm{F}}^{\ast} e^{-B\tilde{t}_{\mathrm{F}}^{\ast}} - \frac{B\tilde{t}_{\mathrm{F}}^{\ast}}{2} e^{-\frac{B\tilde{t}_{\mathrm{F}}^{\ast}}{2}} \right) < 0$ since $\frac{B\tilde{t}_{\mathrm{F}}^{\ast}}{2} \geq 1$ can be satisfied. The inequality (c) holds due to the inequality $\sqrt{xy} \leq \frac{x+y}{2}$. In addition, the equality in (a) holds if $\{\tilde{\bm{\rho}}_{\mathrm{F}}^{\ast}, \tilde{t}_{\mathrm{F}}^{\ast},\tilde{d}_{\mathrm{F}}^{\ast}\}$ is also the optimal solution to the problem (NOMA-F); the equality in (b) holds if $\tilde{h}_{\mathrm{F},1}^{\ast} = \tilde{h}_{\mathrm{F},2}^{\ast}$; the equality in (c) holds if $\bar{\Upsilon}_1^{\mathrm{I}}(\tilde{\rho}_{\mathrm{F},1}^{\ast},\frac{\tilde{t}_{\mathrm{F}}^{\ast}}{2}) = \bar{\Upsilon}_2^{\mathrm{I}}(\tilde{\rho}_{\mathrm{F},2}^{\ast},\frac{\tilde{t}_{\mathrm{F}}^{\ast}}{2})$. Note that the similar result can be also proved if $h_1 < h_2$. We can find that NOMA and FDMA offloadings have the same computation-related energy consumption, i.e., $\{E_k^{\mathrm{loc}}\}$ and $\{E_k^{\mathrm{rem}}\}$, with any given $\{\bm{\rho},t\}$, so the inequality relationship in \eqref{deqn_ex2h} implies that NOMA can achieve a lower MEC-related energy consumption than FDMA with infinite blocklength due to the spectrum resource sharing and the perfect SIC.

\subsection{Comparison between NOMA and FDMA Offloadings with Finite Blocklength}
Then, we compare the MEC-related energy consumption of NOMA and FDMA offloadings with finite blocklength. For a special case that $\eta = \frac{1}{2}$, let $\{\bm{\rho}_{\mathrm{N}}^{\ast}, t_{\mathrm{N}}^{\ast},d_{\mathrm{N}}^{\ast}\}$ denote the optimal solution to problem (NOMA-F), and $\{\bar{h}_{\mathrm{N},k}^{\ast}\}$ and $N_{\mathrm{N}}^{\ast}$ denote the corresponding normalized channel power gains and blocklength, respectively, the energy gap between NOMA and FDMA, i.e., the difference between \eqref{deqn_ex9c} and \eqref{deqn_ex4e}, can be expressed as
\begin{align}
  & E^{\mathrm{NOMA}}_{\mathrm{F}}(\bm{\rho},t,d) - E^{\mathrm{FDMA}}_{\mathrm{F}}(\bm{\rho},t,d) \nonumber\\
  \overset{\text{(a)}}{\geq} & E^{\mathrm{NOMA}}_{\mathrm{F}}(\bm{\rho}_{\mathrm{N}}^{\ast}, t_{\mathrm{N}}^{\ast},d_{\mathrm{N}}^{\ast}) - E^{\mathrm{FDMA}}_{\mathrm{F}}(\bm{\rho}_{\mathrm{N}}^{\ast}, t_{\mathrm{N}}^{\ast},d_{\mathrm{N}}^{\ast}) \nonumber\\
  \overset{\text{(b)}}{>} & \left[ \frac{\Upsilon_1^{\mathrm{F}}(\rho_{\mathrm{N,1}}^{\ast},N_{\mathrm{N}}^{\ast}) (\Upsilon_2^{\mathrm{F}}(\rho_{\mathrm{N,2}}^{\ast},N_{\mathrm{N}}^{\ast}) + 1)}{\bar{h}_{\mathrm{N},1}^{\ast}} + \frac{\Upsilon_2^{\mathrm{F}}(\rho_{\mathrm{N,2}}^{\ast},N_{\mathrm{N}}^{\ast})}{\bar{h}_{\mathrm{N},2}^{\ast}} \right] \frac{N_{\mathrm{N}}^{\ast}}{B} \nonumber\\
  & \quad - \left[ \frac{\Upsilon_1^{\mathrm{F}}(\rho_{\mathrm{N,1}}^{\ast},\frac{N_{\mathrm{N}}^{\ast}}{2})}{2\bar{h}_{\mathrm{N},1}^{\ast}} + \frac{\Upsilon_2^{\mathrm{F}}(\rho_{\mathrm{N,2}}^{\ast},\frac{N_{\mathrm{N}}^{\ast}}{2})}{2\bar{h}_{\mathrm{N},2}^{\ast}} \right] \frac{N_{\mathrm{N}}^{\ast}}{B} \nonumber\\
  = & \left[\frac{\bar{\Upsilon}_1^{\mathrm{F}}(\rho_{\mathrm{N,1}}^{\ast},N_{\mathrm{N}}^{\ast}) \bar{\Upsilon}_2^{\mathrm{F}}(\rho_{\mathrm{N,2}}^{\ast},N_{\mathrm{N}}^{\ast})}{\bar{h}_{\mathrm{N},1}^{\ast}} - \sum_{k=1}^{2}{\frac{\bar{\Upsilon}_k^{\mathrm{F}}(\rho_{\mathrm{N},k}^{\ast},\frac{N_{\mathrm{N}}^{\ast}}{2})}{2 \bar{h}_{\mathrm{N},k}^{\ast}}} \right. \nonumber\\
  & \quad \left. + \left(\frac{1}{\bar{h}_{\mathrm{N},2}^{\ast}} - \frac{1}{\bar{h}_{\mathrm{N},1}^{\ast}}\right) \left(\bar{\Upsilon}_2^{\mathrm{F}}(\rho_{\mathrm{N,2}}^{\ast},N_{\mathrm{N}}^{\ast}) + \frac{1}{2}\right) \right] \frac{N_{\mathrm{N}}^{\ast}}{B} \triangleq \bigtriangleup_{\mathrm{F}}^{\mathrm{N}}, \label{deqn_ex3h}
\end{align}
where $\bar{\Upsilon}_k^{\mathrm{F}}(\rho_k,N) \triangleq \Upsilon_k^{\mathrm{F}}(\rho_k,N) + 1 = \mathrm{exp} ( \frac{\mathrm{ln}2 \rho_k L_k}{N(1 - \epsilon_k)} + \frac{Q^{-1}(\epsilon_k)}{\sqrt{N}} )$. In \eqref{deqn_ex3h}, the equality in (a) holds if $\{\bm{\rho}_{\mathrm{N}}^{\ast}, t_{\mathrm{N}}^{\ast},d_{\mathrm{N}}^{\ast}\}$ is also the optimal solution to problem (FDMA-F), and the inequality (b) holds because $\check{P}_{\mathrm{F},k}^{\mathrm{NOMA}} > \hat{P}_{\mathrm{F},k}^{\mathrm{NOMA}}, \forall k$. Since $\bar{\Upsilon}_k^{\mathrm{F}}(\rho_k,N)$ is complex and the value of \eqref{deqn_ex3h} depends on the relative magnitude between $\rho_{\mathrm{N},1}^{\ast}$ and $\rho_{\mathrm{N},2}^{\ast}$ and that between $\bar{h}_{\mathrm{N},1}^{\ast}$ and $\bar{h}_{\mathrm{N},2}^{\ast}$, it is hard to determine whether $\bigtriangleup_{\mathrm{F}}^{\mathrm{N}}$ is positive or negative directly.

\begin{figure}[!t]
  \centering
  \subfloat[$\mathcal{A}(\rho_1,\rho_2,N)$]{
      \hspace{-0.2cm}\includegraphics[width=1.8in]{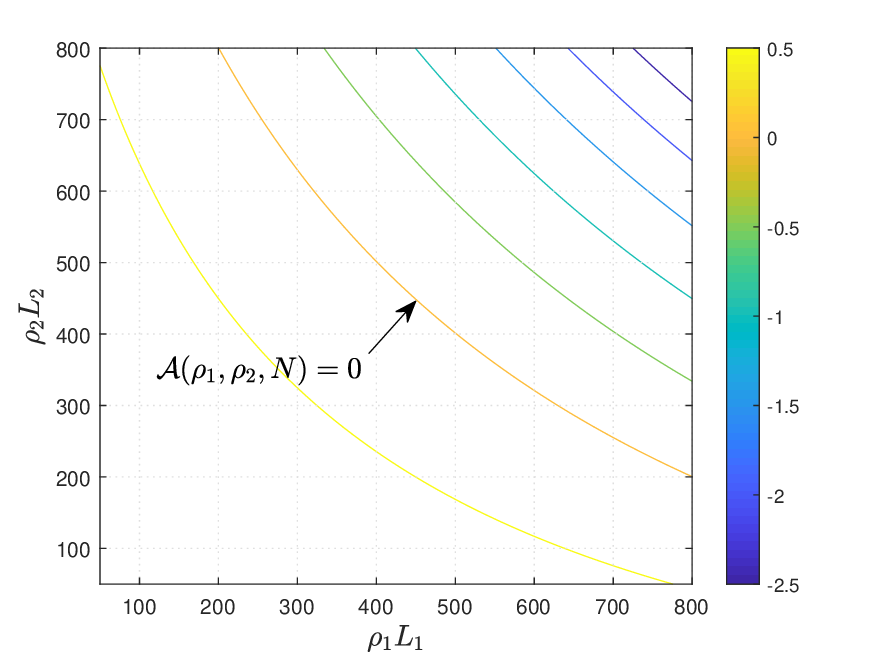}}
  \subfloat[$\mathcal{B}(\rho_1,\rho_2,N)$]{
      \includegraphics[width=1.8in]{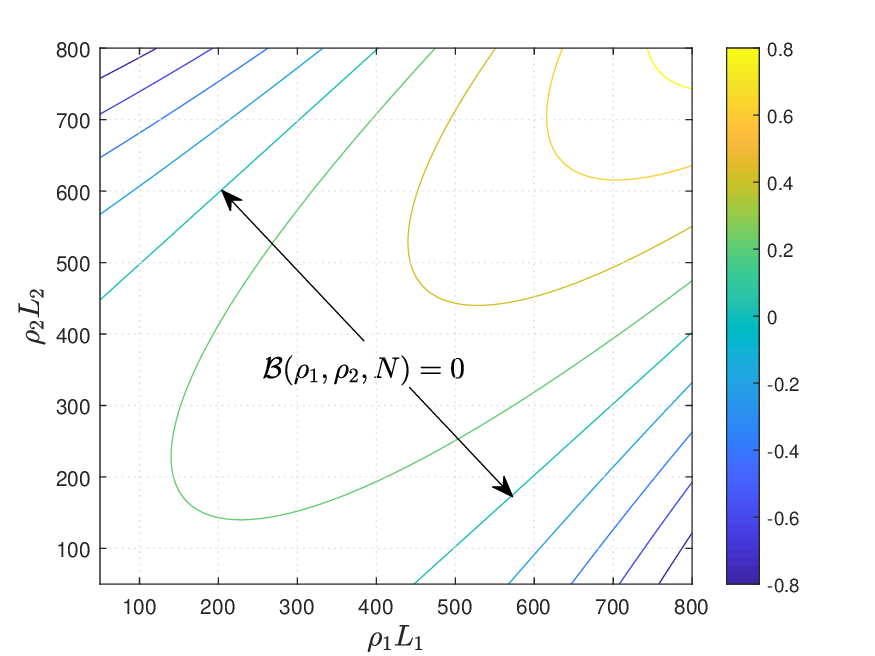}}
  \caption{Isolines of $\mathcal{A}(\rho_1,\rho_2,N)$ and $\mathcal{B}(\rho_1,\rho_2,N)$ over mesh $(\rho_1L_1,\rho_2L_2)$ for $N = 300$.}
  \label{fig_isoline}
\end{figure}

In order to facilitate the analysis of the value of $\bigtriangleup_{\mathrm{F}}^{\mathrm{N}}$ within its feasible domain for a given $N$, we first define $\mathcal{A}(\rho_1,\rho_2,N) \triangleq 1 - \Upsilon_1^{\mathrm{F}}(\rho_1,N) \Upsilon_2^{\mathrm{F}}(\rho_2,N)$ and $\mathcal{B}(\rho_1,\rho_2,N) \triangleq \bar{\Upsilon}_1^{\mathrm{F}}(\rho_1,N) \bar{\Upsilon}_2^{\mathrm{F}}(\rho_2,N) - \frac{1}{2}\sum_{k=1}^{2}{\bar{\Upsilon}_k^{\mathrm{F}}(\rho_k,\frac{N}{2})}$, and show their isolines over mesh $(\rho_1L_1,\rho_2L_2)$ for $N$ = 300 in Fig. \ref{fig_isoline}. In Fig. \ref{fig_isoline}(a), we can observe that $\mathcal{A}(\rho_1,\rho_2,N)$ decreases with the increase of $\rho_1L_1 + \rho_2L_2$ and $\mathcal{A}(\rho_1,\rho_2,N) > 0$ should be satisfied to ensure $\check{P}_{\mathrm{F},k}^{\mathrm{NOMA}} \geq 0, \forall k$, since for the given SNR $\gamma_{\mathrm{F},k}^{\mathrm{NOMA}}$ and blocklength $N$, the decoding error probability $\epsilon_k$ increases as the data rate $\frac{\rho_k L_k}{N}$ increases \cite{ref_Sun}, the sum data size of two UEs' offloaded tasks, i.e., $\rho_1L_1 + \rho_2L_2$, cannot exceed a certain value that satisfies $\mathcal{A}(\rho_1,\rho_2,N) = 0$ to ensure the decoding error probabilities of two UEs' offloaded tasks do not exceed $\epsilon_k$. In Fig. \ref{fig_isoline}(b), we can find that $\mathcal{B}(\rho_1,\rho_2,N)$ decreases with the increase of $|\rho_1L_1 - \rho_2L_2 |$. Since $\bigtriangleup_{\mathrm{F}}^{\mathrm{N}} = \frac{N_{\mathrm{N}}^{\ast}}{\bar{h}_{\mathrm{N},1}^{\ast}B} \mathcal{B}(\rho_{\mathrm{N},1}^{\ast},\rho_{\mathrm{N},2}^{\ast},N_{\mathrm{N}}^{\ast})$ when $\bar{h}_{\mathrm{N},1}^{\ast} = \bar{h}_{\mathrm{N},2}^{\ast}$, and $\bar{P}_{\mathrm{F},k}^{\mathrm{NOMA}}$ in \eqref{deqn_ex8c} increases as $\rho_k$ increases, NOMA offloading should keeps an offloaded task gap between two UEs, i.e., $|\rho_1L_1 - \rho_2L_2 |$ to leverage the advantages of spectrum efficiency due to the power domain multiplexing.

Thus, if $(\rho_{\mathrm{N,1}}^{\ast}L_1,\rho_{\mathrm{N,2}}^{\ast}L_2)$ is at the overlap area of $\mathcal{A}(\rho_1,\rho_2,N) > 0$ in Fig. \ref{fig_isoline}(a) and $\mathcal{B}(\rho_1,\rho_2,N) \geq 0$ in Fig. \ref{fig_isoline}(b), $\bigtriangleup_{\mathrm{F}}^{\mathrm{N}} \geq 0$ when $\bar{h}_{\mathrm{N},1}^{\ast} = \bar{h}_{\mathrm{N},2}^{\ast}$. For the most special case that $\bar{h}_{\mathrm{N},1}^{\ast} = \bar{h}_{\mathrm{N},2}^{\ast}$, $\rho_{\mathrm{N,1}}^{\ast}L_1 = \rho_{\mathrm{N,2}}^{\ast}L_2$, and $\epsilon_1 = \epsilon_2$, we have
\begin{align}
  \bigtriangleup_{\mathrm{F}}^{\mathrm{N}} = & \frac{N_{\mathrm{N}}^{\ast}}{\bar{h}_{\mathrm{N},k}^{\ast}B} \left( {\bar{\Upsilon}_k^{\mathrm{F}}(\rho_{\mathrm{N},k}^{\ast},N_{\mathrm{N}}^{\ast})}^2 - \bar{\Upsilon}_k^{\mathrm{F}}(\rho_{\mathrm{N},k}^{\ast},\frac{N_{\mathrm{N}}^{\ast}}{2}) \right) \nonumber\\
  > & 0, \; k = 1 \; \mathrm{or} \; 2.\label{deqn_ex4h}
\end{align}
Note that because \eqref{deqn_ex3h} is a strict inequality, and the inequality (b) in \eqref{deqn_ex3h} makes $\bigtriangleup_{\mathrm{F}}^{\mathrm{N}}$ only measures the difference between the MEC-related energy consumption required by SIC success case of NOMA offloading and that required by FDMA offloading, so the above result is conservative.

\begin{figure*}[!t]
	\centering
	\subfloat[Infinite blocklength case]{
		\includegraphics[width=3in]{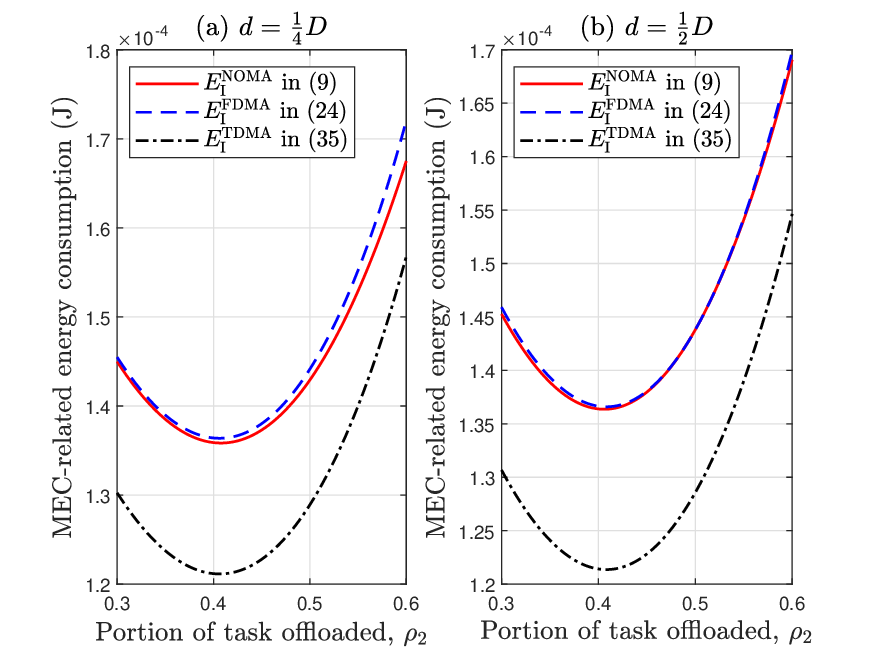}}
	\hspace{+1cm}\subfloat[Finite blocklength case]{
		\includegraphics[width=3in]{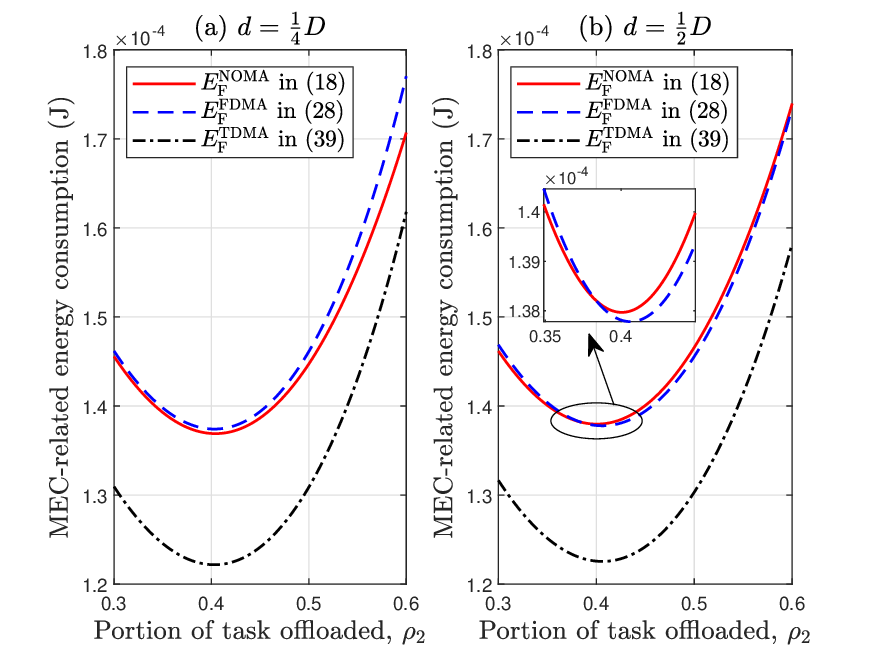}}
	\caption{MEC-related energy consumption required by different multiple access offloading schemes versus the portion of task offloaded by UE~2 $\rho_2$ for different values of $d$ when $\eta = \delta = 0.5$, $D = 100$~m, $\rho_1 = 0.5$, $B = 0.5$~MHz, $t = 0.3$~ms, $L_1 = L_2 = 1.2$~kbit, and $\epsilon_1 = \epsilon_2 = 10^{-5}$.}
	\label{fig_comparison_task}
\end{figure*}

The correctness of the above theoretical conclusion can be verified through the numerical result shown in Fig. \ref{fig_comparison_task}, which presents the calculation results of the MEC-related energy consumption required by different multiple access offloading schemes in both the infinite and finite blocklength cases versus the portion of task offloaded by UE~2 $\rho_2$ for $\rho_1 = 0.5$. As the comparison result in \eqref{deqn_ex1h}, TDMA achieves a lower energy consumption than FDMA in all cases in Fig. \ref{fig_comparison_task}. As the comparison result in \eqref{deqn_ex2h}, if the blocklength is assumed to be infinite, NOMA requires less energy consumption than FDMA, and the performance advantage of NOMA when $d = \frac{1}{4}D$ is more obvious than that when $d = \frac{1}{2}D$ due to the quality difference of the offloading links caused by the UAV location. For the finite-blocklength offloading, NOMA can still achieve a lower energy consumption than FDMA when $d = \frac{1}{4}D$ as shown in Fig. \ref{fig_comparison_task}(a), while the performance advantage of NOMA over FDMA cannot be obtained when 0.4 $\leq \rho_2 \leq$ 0.58 and $d = \frac{1}{2}D$, i.e., $\bar{h}_1 = \bar{h}_2$, in Fig. \ref{fig_comparison_task}(b), which is consistent with the analysis results in Section \uppercase\expandafter{\romannumeral3}-C.

\begin{figure*}[!t]
	\centering
	\subfloat[Infinite blocklength case]{
		\includegraphics[width=3in]{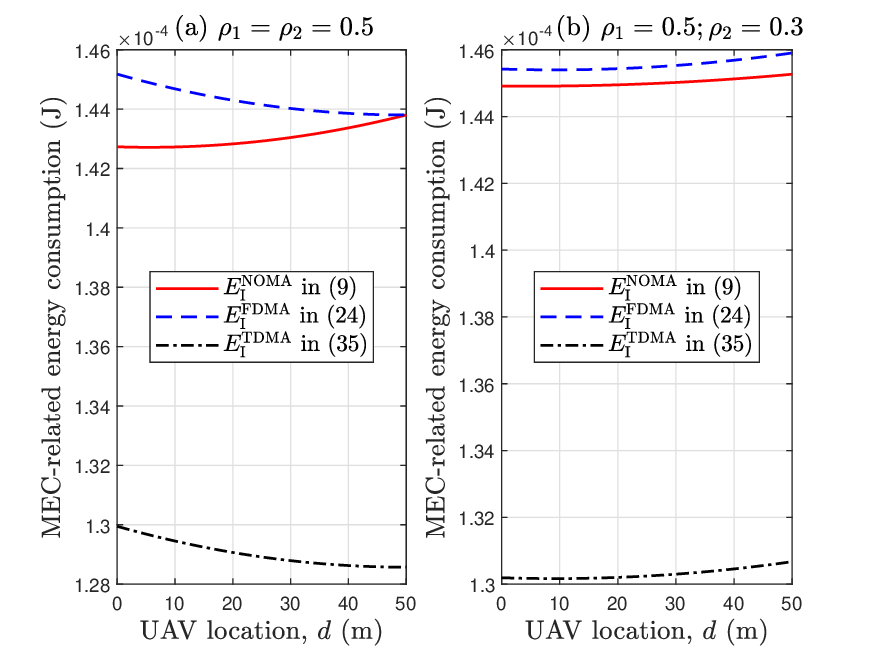}}
	\hspace{+1cm}\subfloat[Finite blocklength case]{
		\includegraphics[width=3in]{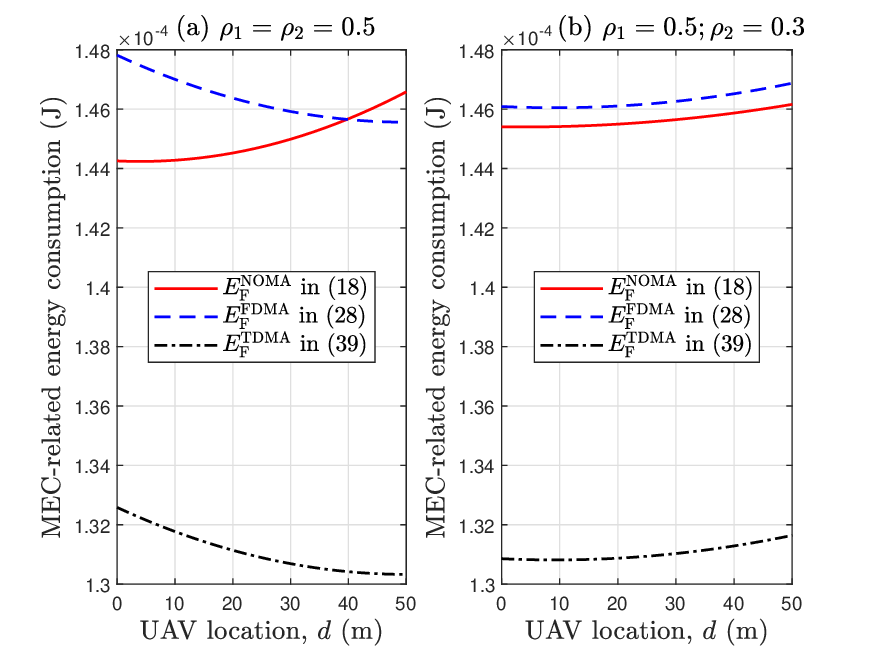}}
	\caption{MEC-related energy consumption required by different multiple access offloading schemes versus the UAV location $d$ for different cases of $\bm{\rho}$ when $\eta = \delta = 0.5$, $D = 100$~m, $B = 0.5$~MHz, $t = 0.3$~ms, $L_1 = L_2 = 1.2$~kbit, and $\epsilon_1 = \epsilon_2 = 10^{-5}$.}
	\label{fig_comparison_UAV}
\end{figure*}

To investigate the impact of UAV location on the performance gap between NOMA and FDMA, we present the MEC-related energy consumption versus UAV location $d$ in Fig.~\ref{fig_comparison_UAV}. For the symmetric task offloading case (i.e., $\rho_1 = \rho_2 = 0.5$, where the computation-related energy consumption of two UEs is identical), as $d$ decreases from $\frac{D}{2} = 50$~m to 0~m, the energy consumption gap between NOMA and FDMA increases, as shown in both Fig.~\ref{fig_comparison_UAV}(a) and Fig.~\ref{fig_comparison_UAV}(b). This is because the performance benefits of NOMA over FDMA become more pronounced when the channel conditions of the users differ significantly~\cite{ref_NOMA_chan}. When the offloading channels of two UEs tend to be symmetric (i.e., $d = \frac{D}{2} = 50$~m), the performance gap becomes zero if the blocklength is infinite. However, in the finite blocklength case, NOMA requires more energy consumption than FDMA due to the consideration of decoding error probabilities and SIC failure. In contrast, for the asymmetric task offloading case (i.e., $\rho_1 = 0.5$ and $\rho_2 = 0.3$), NOMA consistently achieves lower energy consumption compared to FDMA with both infinite and finite blocklength, thanks to the efficiency provided by the power domain multiplexing.

To sum up, TDMA can achieve a lower MEC-related energy consumption than FDMA in both infinite blocklength and finite blocklength cases due to the sequential task offloading of the UEs, and NOMA consumes less energy compared to FDMA in infinite blocklength case. However, the same conclusion cannot be obtained if the blocklength is finite, especially when the channel power gains and the offloaded task data sizes of two UEs are relatively symmetrical, the energy consumption of NOMA will be larger than that of FDMA.

\section{Proposed Algorithm for Problem (NOMA-F)}
Note that the six MEC-related energy consumption minimization problems of different offloading schemes formulated in Section \uppercase\expandafter{\romannumeral2} are all non-convex due to the non-convexity of their objective functions and constraints, as well as the coupling of the optimization variables, especially in the expressions of transmit power, which make these problems difficult to be solved directly. We can find that the mathematical forms of these six problems are similar except for the specific expressions of the transmit power. To avoid redundancy, we only present the solution to problem (NOMA-F) in the following, which is the most complex and challenging, and the other five problems can be solved similarly. To overcome the above difficulties, we decouple variables $\bm{\rho}$, $t$, and $d$ in problem (NOMA-F) by the BCD technique, which decomposes problem (NOMA-F) into three subproblems: 1) Subproblem 1 optimizes the portions of task offloaded $\bm{\rho}$ for given $t$ and $d$; 2) Subproblem 2 optimizes the offloading time $t$ for given $\bm{\rho}$ and $d$; 3) Subproblem 3 optimizes the UAV location $d$ for given $\bm{\rho}$ and $d$. These three subproblems are solved in an alternating manner until the objective value $E_{\mathrm{F}}^{\mathrm{NOMA}}(\bm{\rho},t,d)$ converges. The solutions to the three subproblems are presented as follows.

\subsection{Optimizing the Portions of Task Offloaded}
For given $d$ and $t$, problem (NOMA-F) reduces to
\begin{subequations}\label{eq:7}
  \begin{alignat}{2}
    \min_{\bm{\rho}} \quad & \sum_{k=1}^{2}{\zeta_k \rho_k^3 + \xi_k (1-\rho_k)^3 + \bar{P}_{\mathrm{F},k}^{\mathrm{NOMA}} t} & \label{eq:7A}\\
    \mbox{s.t.} \quad 
    & \eqref{eq:1B}, \eqref{eq:1C}, \eqref{eq:1E},\eqref{eq:2B},\eqref{eq:2C}, & \label{eq:7B}
  \end{alignat}
\end{subequations}
where $\zeta_k = \frac{\kappa_{\mathrm{U}} (c_k L_k)^3}{(T_{\mathrm{max}} - t)^2}$ and $\xi_k = \frac{\kappa_k (c_k L_k)^3}{{T_{\mathrm{max}}}^2}$. The expressions of transmit power in \eqref{deqn_ex6c} and \eqref{deqn_ex7c} can be simplified as
\begin{subequations}\label{deqn_ex1i}
  \begin{align}
   \hat{P}_{\mathrm{F},1}^{\mathrm{NOMA}}(\bm{\rho}) &= \frac{1}{\bar{h}_1} \left( e^{\omega_1\rho_1 + \omega_2\rho_2 + v_1 + v_2} - e^{\omega_2\rho_2 + v_2} \right), \label{deqn_ex1i_A}\\
   \hat{P}_{\mathrm{F},2}^{\mathrm{NOMA}}(\bm{\rho}) &= \frac{1}{\bar{h}_2} \left( e^{\omega_2\rho_2 + v_2} - 1 \right), \label{deqn_ex1i_B}\\
   \check{P}_{\mathrm{F},1}^{\mathrm{NOMA}}(\bm{\rho}) &= \frac{1}{\bar{h}_1} \left( \frac{1}{1 - e^{\omega_2 \rho_2}(e^{v_2} - e^{-\omega_1 \rho_1})} - 1 \right), \label{deqn_ex1i_C}\\
   \check{P}_{\mathrm{F},2}^{\mathrm{NOMA}}(\bm{\rho}) &= \frac{1}{\bar{h}_2} \left( \frac{1}{1 - e^{\omega_1 \rho_1}(e^{v_1} - e^{-\omega_2 \rho_2})} - 1 \right), \label{deqn_ex1i_D}
  \end{align}
\end{subequations}
where $\omega_k \triangleq \frac{\mathrm{ln}2 L_k}{N(1-\epsilon_k)}$ and $v_k \triangleq \frac{Q^{-1}(\epsilon_k)}{\sqrt{N}}$. To handle the complex forms of $\{\check{P}_{\mathrm{F},k}^{\mathrm{NOMA}}(\bm{\rho})\}$, we introduce slack variables $\tilde{\bm{\lambda}} \triangleq \{\tilde{\lambda}_1,\tilde{\lambda}_2\}$, and $\tilde{\bm{\mu}} \triangleq \{\tilde{\mu}_1,\tilde{\mu}_2\}$, and problem \eqref{eq:7} can be equivalently transformed into
\begin{subequations}\label{eq:8}
  \begin{alignat}{2}
    \min_{\bm{\rho},\tilde{\bm{\lambda}},\tilde{\bm{\mu}}} \quad & \sum_{k=1}^{2}{\zeta_k \rho_k^3 + \xi_k (1-\rho_k)^3} + \mathcal{P}_k(\bm{\rho},\tilde{\mu}_k)t & \label{eq:8A}\\
    \mbox{s.t.} \quad 
    & \tilde{\mu}_k \leq 1 - e^{\tilde{\lambda}_k}, \; \forall k, & \label{eq:8B}\\
    & 0 \geq \tilde{\lambda}_k \geq \lambda_k(\bm{\rho}), \; \forall k, & \label{eq:8C}\\
    & \mathcal{P}_k(\bm{\rho},\tilde{\mu}_k) \leq P_{k,\mathrm{max}}, \; \forall k, & \label{eq:8D}\\
    & \eqref{eq:1B}, \eqref{eq:1C}, \eqref{eq:1E}, & \label{eq:8E}
  \end{alignat}
\end{subequations}
where $\mathcal{P}_k(\bm{\rho},\tilde{\mu}_k) \triangleq (1 - \epsilon_1)\hat{P}_{\mathrm{F},k}^{\mathrm{NOMA}}(\bm{\rho}) + \frac{\epsilon_1}{\bar{h}_k}\left( \frac{1}{\tilde{\mu}_k} - 1 \right)$, $\lambda_1(\bm{\rho}) \triangleq \mathrm{ln}\left( e^{\omega_2 \rho_2}(e^{v_2} - e^{-\omega_1 \rho_1}) \right)$, and $\lambda_2(\bm{\rho}) \triangleq \mathrm{ln}\left( e^{\omega_1 \rho_1}(e^{v_1} - e^{-\omega_2 \rho_2}) \right)$. Note that $\tilde{\lambda}_k \leq 0$ in \eqref{eq:8C} is equivalent to \eqref{eq:2C} to guarante that $\check{P}_{\mathrm{F},k}^{\mathrm{NOMA}}(\bm{\rho}) \geq 0, \forall k$. Problem \eqref{eq:8} is still non-convex due to the non-convexity of $\hat{P}_{\mathrm{F},1}^{\mathrm{NOMA}}(\bm{\rho})$ in \eqref{eq:8A} and $\{\lambda_k(\bm{\rho})\}$ in \eqref{eq:8C}, which all contain a difference of two convex functions. This form can be effectively handled by the SCA technique \cite{ref_DC} that replaces the non-convex objective function \eqref{eq:8A} and constraint \eqref{eq:8C} with suitable convex approximations, e.g., approximating $e^{\omega_2\rho_2 + v_2}$ in $\hat{P}_{\mathrm{F},1}^{\mathrm{NOMA}}(\bm{\rho})$ and $e^{-\omega_k \rho_k}$ in $\lambda_k(\bm{\rho})$ in the $i$th iteration with their first-order Taylor expansion at the given local point $\rho_k^{(i-1)}$ that denotes the solution of $\rho_k$ obtained in the $(i-1)$th iteration, and solves problem \eqref{eq:8} iteratively until its objective value converges.

\subsection{Optimizing the Offloading Time}
For given $\bm{\rho}$ and $d$, problem (NOMA-F) reduces to
\begin{subequations}\label{eq:10}
  \begin{alignat}{2}
    \min_{t} \quad & \sum_{k=1}^{2}{\kappa_{\mathrm{U}}\frac{(\rho_k c_k L_k)^3}{(T_{\mathrm{max}} - t)^2} + \bar{P}_{\mathrm{F},k}^{\mathrm{NOMA}} t} & \label{eq:10A}\\
    \mbox{s.t.} \quad 
    & \eqref{eq:1B}, \eqref{eq:1F}, \eqref{eq:2B}, \eqref{eq:2C}. & \label{eq:10B}
  \end{alignat}
\end{subequations}
The expressions of transmit power in \eqref{deqn_ex6c} and \eqref{deqn_ex7c} can be simplified as
\begin{subequations}\label{deqn_ex3i}
  \begin{align}
   \hat{P}_{\mathrm{F},1}^{\mathrm{NOMA}}(t) &= \frac{1}{\bar{h}_1} \left( e^{\frac{z_1+z_2}{t} + \frac{s_1+s_2}{\sqrt{t}}} - e^{\frac{z_2}{t} + \frac{s_2}{\sqrt{t}}} \right), \label{deqn_ex3i_A}\\
   \hat{P}_{\mathrm{F},2}^{\mathrm{NOMA}}(t) &= \frac{1}{\bar{h}_2} \left( e^{\frac{z_2}{t} + \frac{s_2}{\sqrt{t}}} - 1 \right), \label{deqn_ex3i_B}\\
   \check{P}_{\mathrm{F},1}^{\mathrm{NOMA}}(t) &= \frac{1}{\bar{h}_1} \left( \frac{1}{1 - e^{\frac{z_2}{t} + \frac{s_2}{\sqrt{t}}} + e^{\frac{z_2-z_1}{t}}} - 1 \right), \label{deqn_ex3i_C}\\
   \check{P}_{\mathrm{F},2}^{\mathrm{NOMA}}(t) &= \frac{1}{\bar{h}_2} \left( \frac{1}{1 - e^{\frac{z_1}{t} + \frac{s_1}{\sqrt{t}}} + e^{\frac{z_1-z_2}{t}}} - 1 \right), \label{deqn_ex3i_D}
  \end{align}
\end{subequations}
where $z_k = \frac{\mathrm{ln}2 \rho_k L_k}{B(1-\epsilon_k)}$ and $s_k = \frac{Q^{-1}(\epsilon_k)}{\sqrt{B}}$. Similarly, to handle the complex forms of $\{\check{P}_{\mathrm{F},k}^{\mathrm{NOMA}}(t)\}$, by introducing slack variables $\breve{\mathbf{p}} \triangleq \{\breve{p}_1,\breve{p}_2\}$, $\breve{\bm{\varsigma}} \triangleq \{\breve{\varsigma}_1,\breve{\varsigma}_2\}$, and $\breve{\mathbf{e}} \triangleq \{\breve{e}_1,\breve{e}_2\}$, problem \eqref{eq:10} can be equivalently transformed into
\begin{subequations}\label{eq:11}
  \begin{alignat}{2}
    \min_{t,\breve{\mathbf{p}},\breve{\bm{\varsigma}},\breve{\mathbf{e}}} \quad & \sum_{k=1}^{2}{\kappa_{\mathrm{U}}\frac{(\rho_k c_k L_k)^3}{(T_{\mathrm{max}} - t)^2} + \breve{e}_k} & \label{eq:11A}\\
    \mbox{s.t.} \quad 
    & \breve{p}_k \geq \mathcal{G}_k(t,\breve{\varsigma}_k), \; \forall k, & \label{eq:11B}\\
    & \breve{\varsigma}_k \leq \varsigma_k(t), \; \forall k, & \label{eq:11C}\\
    & 0 \leq \breve{\varsigma}_k \leq 1, \; \forall k, & \label{eq:11D}\\
    & \breve{e}_k \geq \breve{p}_k t, \; \forall k, & \label{eq:11E}\\
    & \breve{p}_k \leq P_{k,\mathrm{max}}, \; \forall k, & \label{eq:11F}\\
    & \eqref{eq:1B}, \eqref{eq:1F}, & \label{eq:11G}
  \end{alignat}
\end{subequations}
where $\mathcal{G}_k(t,\breve{\varsigma}_k) \triangleq (1-\epsilon_1)\hat{P}_{\mathrm{F},k}^{\mathrm{NOMA}}(t)+\frac{\epsilon_1}{\bar{h}_k}\left( \frac{1}{\breve{\varsigma}_k} - 1 \right)$, $\varsigma_1(t) \triangleq 1 - e^{\frac{z_2}{t} + \frac{s_2}{\sqrt{t}}} + e^{\frac{z_2-z_1}{t}}$, and $\varsigma_2(t) \triangleq 1 - e^{\frac{z_1}{t} + \frac{s_1}{\sqrt{t}}} + e^{\frac{z_1-z_2}{t}}$, and note that \eqref{eq:11D} is equivalent to \eqref{eq:2C}. Problem \eqref{eq:11} is non-convex due to the non-convexities of constraints \eqref{eq:11B} and \eqref{eq:11C} as well as the bilinear constraints in \eqref{eq:11E}. To overcome this difficulty, we will approximate problem \eqref{eq:11} as a convex one and solve it in an iterative manner.

In the $i$th iteration, $\breve{p}_k t$ in \eqref{eq:11E} can be first rewritten as a difference of two convex functions, i.e., $\breve{p}_k t = \left( \frac{\breve{p}_k + t}{2} \right)^2 - \left( \frac{\breve{p}_k - t}{2} \right)^2$. Then, by applying the first-order Taylor expansion on $(\frac{\breve{p}_k - t}{2})^2$ with the given local point $\{\breve{p}_k^{(i-1)},t^{(i-1)}\}$, where $\breve{p}_k^{(i-1)}$ and $t^{(i-1)}$ respectively denote the solutions of $\breve{p}_k$ and $t$ obtained in the $(i-1)$th iteration, constraint \eqref{eq:11E} can be approximated by
\begin{align}
  \breve{e}_k \geq &\frac{1}{4} (\breve{p}_k + t)^2 + \frac{1}{4} (\breve{p}_k^{(i-1)} - t^{(i-1)})^2 \nonumber\\
  & - \frac{1}{2} (\breve{p}_k^{(i-1)} - t^{(i-1)})(\breve{p}_k - t), \; \forall k. \label{deqn_ex6i}
\end{align}
Finally, the non-convexities of $\hat{P}_{\mathrm{F},1}^{\mathrm{NOMA}}(t)$ in \eqref{eq:11B} and $\{\varsigma_k(t)\}$ in \eqref{eq:11C} that caused by their forms of the difference of two convex functions can be effectively handled by SCA technique, i.e., approximating problem \eqref{eq:11} by replacing $e^{\frac{z_2}{t} + \frac{s_2}{\sqrt{t}}}$ in $\hat{P}_{\mathrm{F},1}^{\mathrm{NOMA}}(t)$, $e^{\frac{z_2-z_1}{t}}$ in $\varsigma_1(t)$, and $e^{\frac{z_1-z_2}{t}}$ in $\varsigma_2(t)$ with their first-order Taylor expansion at the given local points $t^{(i-1)}$.

\subsection{Optimizing the UAV Location}
For given $\bm{\rho}$ and $t$, problem (NOMA-F) reduces to
\begin{subequations}\label{eq:13}
  \begin{alignat}{2}
    \min_{d} \quad & \bar{P}_{\mathrm{F},1}^{\mathrm{NOMA}} + \bar{P}_{\mathrm{F},2}^{\mathrm{NOMA}} & \label{eq:13A}\\
    \mbox{s.t.} \quad 
    & \eqref{eq:1G}, \eqref{eq:2B}. & \label{eq:13B}
  \end{alignat}
\end{subequations}
The objective function in \eqref{eq:13A} is convex with respect to $d$, and the solution of problem \eqref{eq:13} can be derived as
\begin{equation}
  \label{deqn_ex7i}
  d^{\star} = \left[ \frac{\Psi(\rho_1,\rho_2,N)}{\Phi(\rho_1,\rho_2,N) + \Psi(\rho_1,\rho_2,N)}D \right]_{d_{\mathrm{min}}}^{d_{\mathrm{max}}},
\end{equation}
where$[x]_b^a \triangleq \mathrm{min}\{a,\mathrm{max}\{b,x\}\}$, $\Phi(\rho_1,\rho_2,N) \triangleq (1-\epsilon_1)\times \left(\Upsilon_1^{\mathrm{F}}(\rho_1,N) (\Upsilon_2^{\mathrm{F}}(\rho_2,N) +1)\right) + \epsilon_1 \frac{\Upsilon_1^{\mathrm{F}}(\rho_1,N) (\Upsilon_2^{\mathrm{F}}(\rho_2,N) +1)}{1 - \Upsilon_1^{\mathrm{F}}(\rho_1,N) \Upsilon_2^{\mathrm{F}}(\rho_2,N)}$, and $\Psi(\rho_1,\rho_2,N) \triangleq (1-\epsilon_1)\Upsilon_2^{\mathrm{F}}(\rho_2,N) + \epsilon_1 \frac{\Upsilon_2^{\mathrm{F}}(\rho_2,N) (\Upsilon_1^{\mathrm{F}}(\rho_1,N) +1)}{1 - \Upsilon_1^{\mathrm{F}}(\rho_1,N) \Upsilon_2^{\mathrm{F}}(\rho_2,N)}$. In addition, $d_{\mathrm{min}} = D - \sqrt{\frac{P_{2,\mathrm{max}} \beta_0}{BN_0 \Psi(\rho_1,\rho_2,N)} - H^2}$ and $d_{\mathrm{max}} = \mathrm{min} \{ \frac{1}{2}D, \sqrt{\frac{P_{1,\mathrm{max}} \beta_0}{BN_0 \Phi(\rho_1,\rho_2,N)} - H^2} \}$ are obtained from \eqref{eq:1G} and \eqref{eq:2B}.

\begin{algorithm}[!t]
	\caption{Proposed Algorithm for Problem (NOMA-F).}
	\begin{algorithmic}[1] \label{alg1}
		\STATE \textbf{Initialization:} Set initial values for $\bm{\rho}^{(0)}$, $t^{(0)}$, $d^{(0)}$, accuracy $\sigma$, the iteration number $i =$ 0, and calculate $E_{\mathrm{F}}^{\mathrm{NOMA}}(\bm{\rho}^{(0)},t^{(0)},d^{(0)})$ according to \eqref{deqn_ex9c}.
		\REPEAT
		\STATE Set $i=i+1$.
		\STATE Given $\bm{\rho}^{(i-1)}$, $t^{(i-1)}$, and $d^{(i-1)}$, obtain $\bm{\rho}^{(i)}$ by solving problem \eqref{eq:8} with the SCA-based algorithm.
		\STATE Given $\bm{\rho}^{(i)}$, $t^{(i-1)}$, and $d^{(i-1)}$ and calculate $\breve{p}_k^{(i-1)}$ according to \eqref{deqn_ex8c}, obtain $t^{(i)}$ by solving problem \eqref{eq:11} with the SCA-based algorithm.
		\STATE Given $\bm{\rho}^{(i)}$ and $t^{(i)}$, obtain $d^{(i)}$ according to \eqref{deqn_ex7i}.
		\STATE Calculate $E_{\mathrm{F}}^{\mathrm{NOMA}}(\bm{\rho}^{(i)},t^{(i)},d^{(i)})$ according to \eqref{deqn_ex9c}.
		\UNTIL $\left | \frac{E_{\mathrm{F}}^{\mathrm{NOMA}}(\bm{\rho}^{(i)},t^{(i)},d^{(i)}) - E_{\mathrm{F}}^{\mathrm{NOMA}}(\bm{\rho}^{(i-1)},t^{(i-1)},d^{(i-1)})}{E_{\mathrm{F}}^{\mathrm{NOMA}}(\bm{\rho}^{(i-1)},t^{(i-1)},d^{(i-1)})}\right | \le {\sigma}$.
		\STATE Output: $\bm{\rho}^{\ast} = \bm{\rho}^{(i)}$, $t^{\ast} = t^{(i)}$, $d^{\ast} = d^{(i)}$, and $E_{\mathrm{F}}^{\mathrm{NOMA}\ast} = E_{\mathrm{F}}^{\mathrm{NOMA}}(\bm{\rho}^{(i)},t^{(i)},d^{(i)})$.
	\end{algorithmic} 
\end{algorithm}

\subsection{Overall Algorithm}
We summarize the overall algorithm for solving problem (NOMA-F) in Algorithm~\ref{alg1}, where three subproblems will be solved alternatively and iteratively. Since the SCA-based algorithm is guaranteed to converge based on its global convergence and stability~\cite{ref_DC}, in the $i$th iteration, we have $E_{\mathrm{F}}^{\mathrm{NOMA}}(\bm{\rho}^{(i)},t^{(i-1)},d^{(i-1)}) \leq E_{\mathrm{F}}^{\mathrm{NOMA}}(\bm{\rho}^{(i-1)},t^{(i-1)},d^{(i-1)})$ after solving problem \eqref{eq:8}, and $E_{\mathrm{F}}^{\mathrm{NOMA}}(\bm{\rho}^{(i)},t^{(i)},d^{(i-1)}) \leq E_{\mathrm{F}}^{\mathrm{NOMA}}(\bm{\rho}^{(i)},t^{(i-1)},d^{(i-1)})$ after solving problem \eqref{eq:11}. And since $d^{\star}$ in \eqref{deqn_ex7i} is the optimal solution to problem \eqref{eq:13}, we furtherly have $E_{\mathrm{F}}^{\mathrm{NOMA}}(\bm{\rho}^{(i)},t^{(i)},d^{(i)}) \leq E_{\mathrm{F}}^{\mathrm{NOMA}}(\bm{\rho}^{(i-1)},t^{(i-1)},d^{(i-1)})$. Thus, Algorithm~\ref{alg1} is guaranteed to converge since the optimal value is non-increasing and must have a lower bound during each iteration.

The computational complexity of the proposed algorithm mainly depends on the application of CVX solver for solving subproblems that optimize the portions of task offloaded and the offloading time, i.e., problems \eqref{eq:8} and \eqref{eq:11}, in an iterative manner. These two subproblems can be solved by a primal-dual interior point method with complexity $\mathcal{O}(n^{3.5} \mathrm{log}(1/\sigma))$, where $n$ and $\sigma$ denote the number of optimization variables and the given convergence accuracy, respectively~\cite{ref_Convex,ref_Complex}. Since there are $n_1 = 6$ and $n_2 = 5$ optimization variables in problems \eqref{eq:8} and \eqref{eq:11}, respectively, and the computational complexity for calculating the UAV location according to \eqref{deqn_ex7i} can be ignored here, the total computational complexity of Algorithm~\ref{alg1} can be calculated as $\mathcal{O}(I(n_1^{3.5}+n_2^{3.5})\mathrm{log}(1/\sigma))$, where $I$ is the required number of iterations for convergence.\footnote{To mitigate the potential conflict between the computational burden of the optimization algorithm and the strict task computation latency constraints, the optimal computation offloading strategies for various predicted UE locations and task data size ratios can be pre-calculated offline and stored in a database on the network controller.}

\section{Simulation Results}
In this section, we provide some simulation results to compare the performances of different multiple access offloading schemes and demonstrate the effectiveness of the proposed algorithm. The system parameters are set as follows: $N_0 = -169$~dBm/Hz, $\beta_0 = -60$~dB, $D = 100$~m, $H = 50$~m, $f_{\mathrm{U,max}} = 9$~GHz, $\kappa_{\mathrm{U}} = 10^{-28}$, $\eta = \delta = 0.5$, and for both $k \in \{1,2\}$, $f_{k,\mathrm{max}} = 1$~GHz, $P_{k,\mathrm{max}} = 0.1$~W, $c_k = 1000$~cycles/b, $\kappa_k = 10^{-28}$, and $\epsilon_k = \epsilon$, where $\epsilon$ denotes a specific decoding error probability that may take different values in the simulations. The proposed algorithm will be compared with the following three benchmark algorithms for problem (NOMA-F):
\begin{itemize}
  \item{\textit{Without Optimizing the Portions of Task Offloaded:} It fixes the portions of task offloaded as 0.5, i.e., $\rho_k =$ 0.5, $\forall k$, and optimizes the offloading time and the UAV location by solving problems \eqref{eq:11} and \eqref{eq:13}, respectively. The MEC-related energy consumption obtained by this scheme is denoted by $E_{\mathrm{F,NP}}^{\mathrm{NOMA}}$.}
  \item{\textit{Without Optimizing the Offloading Time:} It fixes the offloading time as 0.5$T_{\mathrm{max}}$, and optimizes the portions of task offloaded and the UAV location by solving problems \eqref{eq:8} and \eqref{eq:13}, respectively. The MEC-related energy consumption obtained by this scheme is denoted by $E_{\mathrm{F,NT}}^{\mathrm{NOMA}}$.}
  \item{\textit{Exhaustive Search Method:} It minimizes $E_{\mathrm{F}}^{\mathrm{NOMA}}(\bm{\rho},t,d)$ by searching for the portions of task offloaded and the offloading time in the feasible domain of problem (NOMA-F) and calculating the UAV location according to \eqref{deqn_ex7i}. The MEC-related energy consumption obtained by this scheme is denoted by $E_{\mathrm{F,ES}}^{\mathrm{NOMA}}$, which serves as the globally optimal value of problem (NOMA-F).}
\end{itemize}

\begin{figure}[!t]
	\centering
	\includegraphics[width=3in]{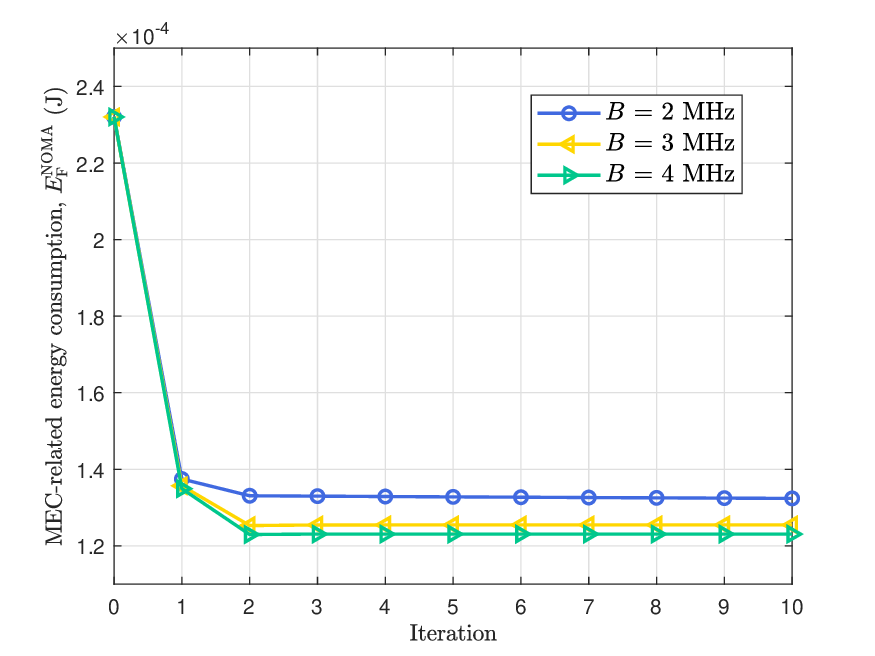}
	\caption{Convergence behavior of the proposed algorithm for problem (NOMA-F) for $B = 2, 3, 4$~MHz.}
	\label{fig_conv}
\end{figure}

First, in Fig. \ref{fig_conv}, for $L_1 = L_2 = 1.2$~kbit, $T_{\mathrm{max}} = 1$~ms, and $\epsilon = 10^{-5}$, we illustrate the convergence behavior of the proposed algorithm (Algorithm \ref{alg1}). It is observed that $E_{\mathrm{F}}^{\mathrm{NOMA}}$ with different $B$ decreases over iterations and converges after three iterations. This indicates that the proposed AO algorithm can converge quickly and demonstrates its effectiveness. Additionally, $E_{\mathrm{F}}^{\mathrm{NOMA}}$ decreases with the increase of $B$ since a larger blocklength $N=Bt$ leads to a higher offloading data rate for each UE. The convergence behaviors of the proposed algorithms for other five offloading schemes are similar, and we do not present them here due to space limitations.

\begin{figure}[!t]
  \centering
  \includegraphics[width=3in]{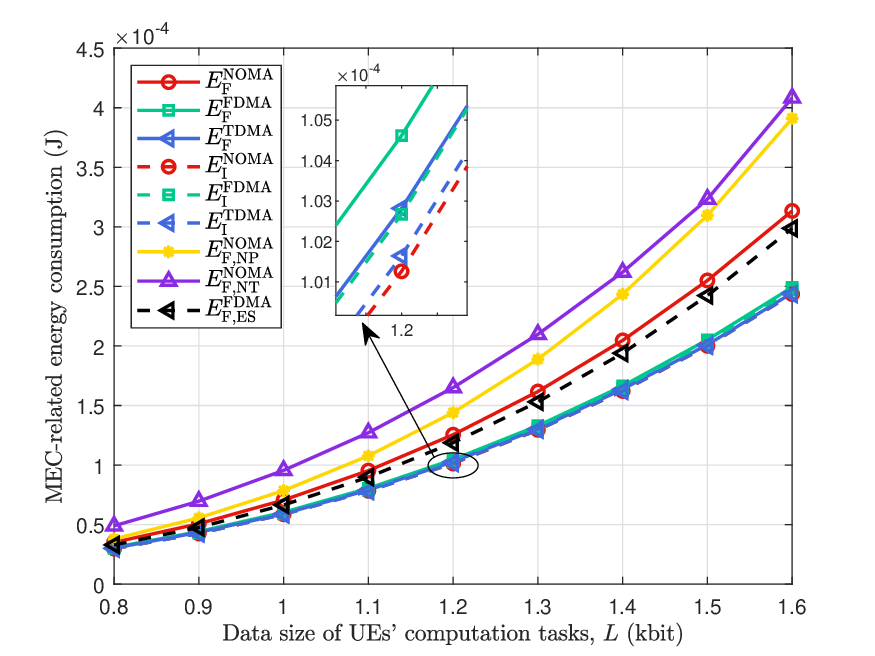}
  \caption{MEC-related energy consumption of different schemes versus data size of UEs' computation tasks $L$ for $B =$ 3 MHz, $T_{\mathrm{max}} =$ 1 ms, and $\epsilon = 10^{-5}$.}
  \label{fig_size}
\end{figure}

In Fig. \ref{fig_size}, by letting $L_1 = L_2 = L$, we show the MEC-related energy consumption of different multiple access offloading schemes and three benchmark schemes versus data size of UEs' computation tasks $L$. As expected, the MEC-related energy consumption of all considered schemes increases as the data size of UEs' computation tasks increases. $E_{\mathrm{F}}^{\mathrm{NOMA}}$ is substantially lower compared with $E_{\mathrm{F,NP}}^{\mathrm{NOMA}}$ and $E_{\mathrm{F,NT}}^{\mathrm{NOMA}}$, and always approaches $E_{\mathrm{F,ES}}^{\mathrm{NOMA}}$, which demonstrates the effectiveness of the proposed joint optimization algorithm. The performance improvement achieved by the proposed algorithm over the first two benchmark schemes increases as $L$ increases, which indicates that the joint optimization of the portions of task offloaded and the offloading time is more important when the computational burden of UEs is heavier.

\begin{figure}[!t]
  \centering
  \includegraphics[width=3in]{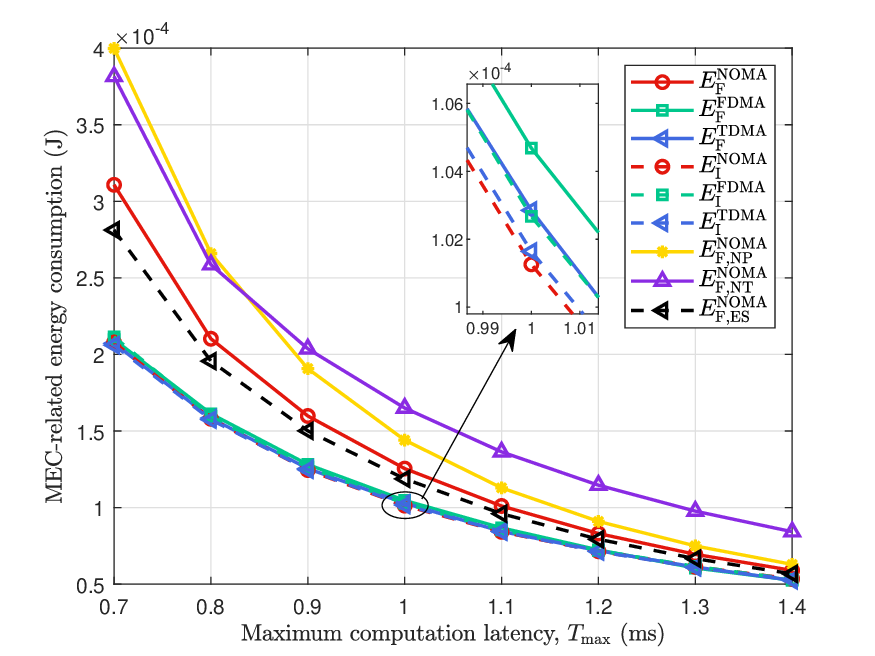}
  \caption{MEC-related energy consumption of different schemes versus maximum computation latency $T_{\mathrm{max}}$ for $L_1 = L_2 =$ 1.2 kbit, $B =$ 3 MHz, and $\epsilon = 10^{-5}$.}
  \label{fig_latency}
\end{figure}

Fig. \ref{fig_latency} shows the MEC-related energy consumption of different schemes versus maximum computation latency $T_{\mathrm{max}}$. It is observed that the MEC-related energy consumption of all schemes decreases with the increase of $T_{\mathrm{max}}$. This is because as the available time for both local computing and remote computing increases, the UEs and UAV can adjust their CPU frequencies to a smaller state according to the principle of the DVFS technique so that the computation-related energy consumption can be effectively reduced. We can also observe that $E_{\mathrm{F}}^{\mathrm{NOMA}}$ is lower than $E_{\mathrm{F,NP}}^{\mathrm{NOMA}}$ and $E_{\mathrm{F,NT}}^{\mathrm{NOMA}}$ regardless of $T_{\mathrm{max}}$. And the gap between $E_{\mathrm{F,NP}}^{\mathrm{NOMA}}$ and $E_{\mathrm{F}}^{\mathrm{NOMA}}$ becomes smaller with the increase of $T_{\mathrm{max}}$, while $E_{\mathrm{F,NT}}^{\mathrm{NOMA}}$ always maintains a larger performance gap with $E_{\mathrm{F}}^{\mathrm{NOMA}}$. This indicates that when the latency requirement is relatively severe, i.e., $T_{\mathrm{max}}$ is small, the joint optimization of the portion of task offloaded and the offloading time is extremely important to reduce energy consumption. Whereas when the latency requirement is relatively loose, i.e., $T_{\mathrm{max}}$ is large, the performance gap caused by the lack of the optimization of the portions of task offloaded can be mitigated by the optimization of offloading time and the moderation effect of the DVFS technique. And the optimization of the offloading time is always important since it should cautiously balance the energy consumption caused by task offloading and remote computing.

In addition, in both Fig. \ref{fig_size} and Fig. \ref{fig_latency}, $E_{\mathrm{F}}^{\mathrm{NOMA}}$ is larger than $E_{\mathrm{F}}^{\mathrm{FDMA}}$ and $E_{\mathrm{F}}^{\mathrm{TDMA}}$ regardless of $L$ and $T_{\mathrm{max}}$. It indicates that the task offloading capability of NOMA is weaker than that of FDMA and TDMA when the blocklength is finite since the SIC failure should be considered additionally for NOMA in this case. Whether in infinite blocklength case or finite blocklength case, TDMA always achieves a lower MEC-related energy consumption compared to FDMA, i.e., $E_{\mathrm{I}}^{\mathrm{TDMA}} < E_{\mathrm{I}}^{\mathrm{FDMA}}$ and $E_{\mathrm{F}}^{\mathrm{TDMA}} < E_{\mathrm{F}}^{\mathrm{FDMA}}$, thanks to the freedom of remote-computing time brought by sequential task offloading of the UEs. Moreover, $E_{\mathrm{I}}^{\mathrm{NOMA}}$ is always lower than $E_{\mathrm{I}}^{\mathrm{FDMA}}$ due to the spectrum efficiency brought by time-frequency resource sharing and the perfect SIC. The above simulation results verify the correctness of the theoretical analysis conclusions obtained in Section \uppercase\expandafter{\romannumeral3}. We can also observe that the MEC-related energy consumption required by the finite blocklength case is always larger than that required by the infinite blocklength case for each considered multiple access offloading scheme because the Shannon capacity services as an upper bound for the finite-blocklength transmission data rate derived in \cite{ref_Poly}.

\begin{figure}[!t]
  \centering
  \includegraphics[width=3in]{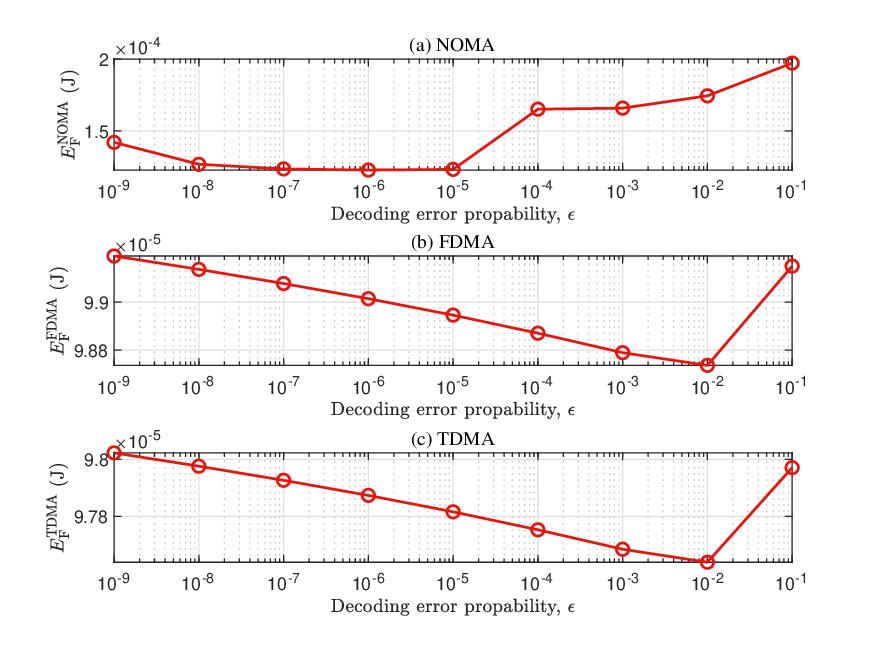}
  \caption{MEC-related energy consumption of different schemes with finite blocklength versus decoding error probability $\epsilon$ for $L_1 = L_2 =$ 1.2 kbit, $B =$ 4 MHz, and $T_{\mathrm{max}} =$ 1 ms.}
  \label{fig_DEP}
\end{figure}

Fig. \ref{fig_DEP} shows the MEC-related energy consumption required by different multiple access offloading schemes with finite blocklength versus decoding error probability $\epsilon$. For any value of $\epsilon$, it can also be observed that $E_{\mathrm{F}}^{\mathrm{NOMA}}$ is larger than $E_{\mathrm{F}}^{\mathrm{FDMA}}$, and $E_{\mathrm{F}}^{\mathrm{FDMA}}$ is larger than $E_{\mathrm{F}}^{\mathrm{TDMA}}$. In Fig. \ref{fig_DEP}(a), it is observed that $E_{\mathrm{F}}^{\mathrm{NOMA}}$ decreases with the increase of $\epsilon$ when $\epsilon \leq 10^{-5}$, and increases with the increase of $\epsilon$ when $\epsilon > 10^{-5}$. This is because $\epsilon$ will affect both the achievable data rate and the SIC failure probability in NOMA offloading as we consider in Section \uppercase\expandafter{\romannumeral2}-A-2. When the reliability requirement is strict, i.e., $\epsilon$ takes an extremely small value, the SIC failure probability is low and the average transmit power derived in \eqref{deqn_ex8c} is dominated by the transmit power required for SIC success case, increasing $\epsilon$ enables the achievable data rate of the offloading link to reach a larger value, which provides a larger computation offloading capacity for UEs to obtain a more flexible task computing strategy and promotes the reduction of the MEC-related energy consumption. However, when the reliability requirement is slack, i.e., $\epsilon$ takes a large value, although increasing $\epsilon$ can still result in a larger achievable data rate, the SIC failure probability is already so large in this case that the transmit power required for SIC failure already occupies a significant portion of the average transmit power, and increasing $\epsilon$ will instead result in a lower effective offloading throughput and a larger communication-related energy consumption due to $\check{P}_{\mathrm{F},k}^{\mathrm{NOMA}} > \hat{P}_{\mathrm{F},k}^{\mathrm{NOMA}}, \; \forall k$. In Fig. \ref{fig_DEP}(b) and Fig. \ref{fig_DEP}(c), as $\epsilon$ increases, both $E_{\mathrm{F}}^{\mathrm{FDMA}}$ and $E_{\mathrm{F}}^{\mathrm{TDMA}}$ decrease slightly when $\epsilon \leq 10^{-2}$ and increase when $\epsilon > 10^{-2}$. Since the UEs offload their tasks to the UAV over orthogonal communication resources in FDMA and TDMA offloadings, the UAV does not need to perform SIC to decode the received UEs' signal, so $\epsilon$ mainly affects the achievable data rate during task offloading transmission, and increasing $\epsilon$ will decrease the energy consumption. However, when $\epsilon$ is too large, i.e., when $\epsilon > 10^{-2}$, the impact of $\epsilon$ on the effective offloading throughput defined in the LHS of \eqref{deqn_ex1e} and \eqref{deqn_ex1g} is significant, increasing $\epsilon$ instead will increase the energy consumption. Based on the above discussion, we can find that the decoding error probability has a more significant impact on NOMA than FDMA and TDMA since both SIC success and failure need to be considered simultaneously when the blocklength is finite.

\begin{figure*}[!t]
  \centering
  \subfloat[Portion of task offloaded]{\includegraphics[width=1.78in]{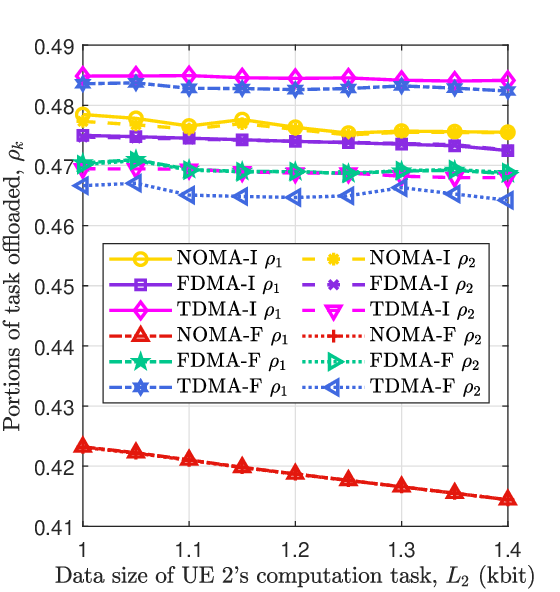}%
  \label{fig_task_ratio}}
  \hfil
  \subfloat[Offloading time]{\includegraphics[width=1.78in]{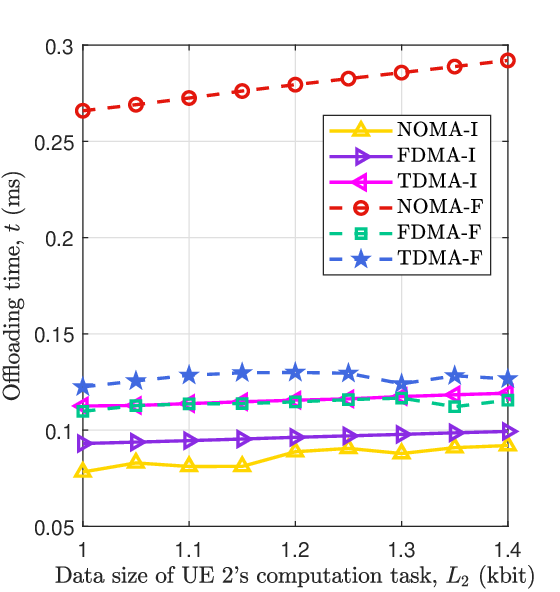}%
  \label{fig_task_time}}
  \hfil
  \subfloat[UAV location]{\includegraphics[width=1.78in]{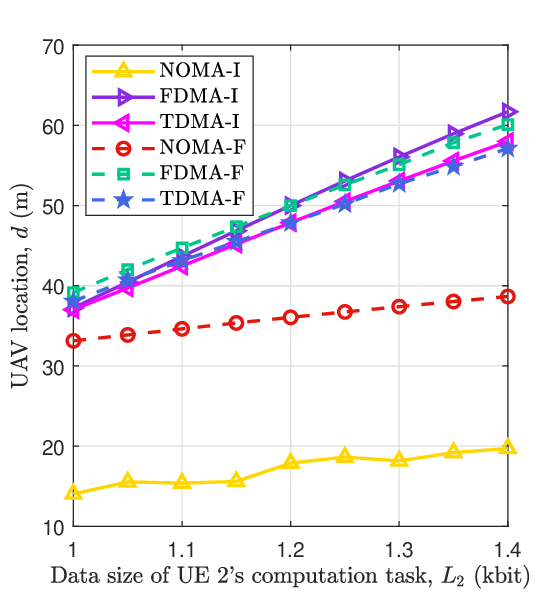}%
  \label{fig_task_loc}}
  \hfil
  \subfloat[Transmit power]{\includegraphics[width=1.78in]{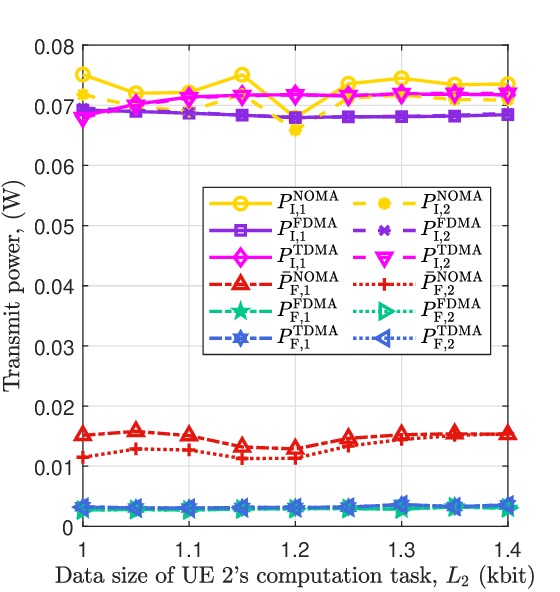}%
  \label{fig_task_power}}
  \caption{Optimized portions of task offloaded, offloading times, UAV locations, and transmit power of different schemes versus data size of UE 2's computation task $L_2$ for $L_1 =$ 1.2 kbit, $B =$ 3 MHz, $T_{\mathrm{max}} =$ 1 ms, and $\epsilon = 10^{-5}$.}
  \label{fig_task}
\end{figure*}

In Fig.~\ref{fig_task}, we show the optimal portions of task offloaded, offloading times, UAV locations, and transmit power of different schemes versus the data size of UE 2's computation task $L_2$ for $L_1 =$ 1.2 kbit. In Fig.~\ref{fig_task}(a) and Fig.~\ref{fig_task}(b), we can observe that the finite-blocklength NOMA offloading scheme obtains the lowest portion of task offloaded and the largest offloading time, which decreases and increases as $L_2$ increases, respectively. This indicates that the finite-blocklength NOMA offloading needs to restrict the offloading throughput and allocate sufficient offloading time to ensure the reliability of offloading transmission can be satisfied even in the SIC failure case as we analyzed in Section \uppercase\expandafter{\romannumeral3}-C. Different from NOMA and FDMA in which $\rho_1$ and $\rho_2$ are very close, $\rho_1$ is higher than $\rho_2$ in both infinite-blocklength and finite-blocklength TDMA offloading schemes since UE 1 offloads its task before UE 2 and then UE 1 can have more time to perform remote computing at the UAV. In Fig. \ref{fig_task}(c), the UAV locations obtained by all schemes become closer to UE 2 as $L_2$ increases to improve the quality of UE 2's offloading link to meet UE 2's task offloading demand. The UAV locations in FDMA and TDMA schemes vary significantly and the ratios of distance from the UAV to UEs 1 and 2 are proportional to the ratios of task data size between UEs 1 and 2. However, in NOMA offloading with both infinite blocklength and finite blocklength, the UAV only moves slightly and it is always closer to UE 1 to ensure $h_1 \geq h_2$ under the assumption that the UAV first decodes UE 1's signal. In Fig. \ref{fig_task}(d), in order to distinguish the signals of two UEs in the power domain, the transmit power of UEs 1 and 2 is different in NOMA schemes, which is different from FDMA and TDMA offloadings. Based on the optimization results presented in Fig. \ref{fig_task}, we can find that whether in infinite blocklength case or finite blocklength case, NOMA mainly balances the difference in offloading demand between two UEs due to the difference between $L_1$ and $L_2$ by adjusting the portions of task offloaded and the offloading time, while FDMA and TDMA mainly balances that by adjusting the UAV locations.

\section{Conclusion}
In this paper, we have considered the minimum MEC-related energy consumption for different multiple access offloading schemes, including NOMA, FDMA, and TDMA in both infinite blocklength and finite blocklength cases. Through theoretical analysis, we have proved that TDMA consistently achieves lower energy consumption than FDMA, and NOMA achieves lower energy consumption than FDMA in the infinite blocklength case. In contrast, since both SIC success and failure cases were considered in finite-blocklength NOMA offloading, NOMA could not be guaranteed to achieve lower energy consumption than FDMA in this case, especially when the channel conditions and the offloaded task data sizes of two UEs are relatively symmetric. Furthermore, we have proposed an alternating optimization algorithm to minimize the MEC-related energy consumption of different schemes by jointly optimizing the portions of task offloaded, the offloading times of all UEs, and the UAV location. Simulation results have validated that the proposed algorithm can effectively reduce MEC-related energy consumption compared to other benchmark schemes and verified the correctness of the theoretical analysis results. In addition, our obtained results have shown that the decoding error probability has a more significant impact on the energy consumption of NOMA than both FDMA and TDMA in the finite blocklength case.

\end{document}